\documentclass[11pt,letterpaper]{article}
\usepackage[margin=1in]{geometry}
\usepackage[T1]{fontenc}
\usepackage{lmodern,amsmath,amssymb,amsthm,microtype,enumitem}
\usepackage{authblk}
\usepackage{tocloft}
\usepackage{tikz,booktabs,xcolor,array}
\usepackage{subcaption}
\usepackage{algorithm}
\usepackage[noend]{algpseudocode}
\usetikzlibrary{arrows.meta,positioning,calc}
\usepackage[colorlinks=true,linkcolor=blue,citecolor=blue,urlcolor=blue]{hyperref}
\hypersetup{pdftitle={Best of Many Worlds for Chores with Sublinear Disposal}}
\newtheorem{theorem}{Theorem}[section]
\newtheorem{lemma}[theorem]{Lemma}
\newtheorem{invariant}[theorem]{Invariant}

\newcommand{\bX}{\mathbf{X}}
\newcommand{\PROP}{\mathsf{PROP}}
\allowdisplaybreaks[1]
\setlist[enumerate]{itemsep=.2em,topsep=.4em}

\newtheorem{definition}[theorem]{Definition}

\title{Best-of-Many-Worlds MMS Allocation for Chores\\
 with Sublinear Disposals}
\author[1]{Zehan Lin}
\author[1]{Xiaowei Wu}
\author[2]{Shengwei Zhou}
\affil[1]{University of Macau, \texttt{\{yc47490,xiaoweiwu\}@um.edu.mo}}
\affil[2]{Nanyang Technological University, \texttt{s.arthur.zhou@gmail.com}}
\date{}

\begin{document}

\hypersetup{pageanchor=false}
\pagestyle{empty}
\maketitle
\thispagestyle{empty}

\begin{abstract}
We study randomized allocations (lotteries) of indivisible chores under additive costs and introduce the notion of MMS1, which requires each agent's bundle to satisfy the maximin share (MMS) bound after the removal of at most one chore.
Our main result establishes that every instance admits an ex-ante proportional lottery in whose support every allocation is MMS1 and all but $o(n)$ agents satisfy MMS.
Consequently, every realized allocation can be made MMS by disposing of at most $o(n)$ chores.
The proposed framework further achieves ex-ante proportionality alongside ex-post $4/3$-MMS, $o(n)$-AMMS, or $1$-out-of-$d$ MMS for $d=(1-o(1))n$.
These guarantees substantially improve upon state-of-the-art bounds for MMS allocation for chores with disposal or subsidy, best-of-both-worlds fairness, approximate AMMS, and ordinal MMS approximations.
\end{abstract}

\definecolor{restoreblue}{RGB}{42,96,150}
\definecolor{restoreorange}{RGB}{215,137,43}
\definecolor{restorered}{RGB}{171,68,68}

\renewcommand{\contentsname}{Contents}
\setcounter{tocdepth}{2}
\setlength{\cftbeforetoctitleskip}{0.6em}
\setlength{\cftaftertoctitleskip}{0.4em}
\setlength{\cftbeforesecskip}{0.3em}
\tableofcontents

\thispagestyle{empty} 
\clearpage
\setcounter{page}{1}  
\pagestyle{plain}
\hypersetup{pageanchor=true}

\section{Introduction}

We study the problem of fairly allocating a set $M$ of $m$ indivisible chores to a group $N$ of $n$ agents with additive cost functions, where each agent evaluates a bundle by the total cost of its chores and prefers bundles of smaller cost.
We focus on the fairness notion of \emph{maximin share} (MMS), introduced by Budish~\cite{journals/jpe/Budish11}.
Informally, the MMS of an agent is the minimum cost she can guarantee herself by partitioning the chores into $n$ bundles and receiving the most costly bundle.
An allocation is MMS if every agent receives a bundle of cost at most her MMS.
Unfortunately, MMS allocations are not guaranteed to exist under additive costs~\cite{conf/aaai/AzizRSW17}.

To address this nonexistence, existing works have pursued three main relaxations.
The first is \emph{multiplicative approximation}, which allows each agent to receive a cost at most $\alpha$ times her MMS, for some $\alpha\ge1$.
For additive chores, \cite{conf/aaai/AzizRSW17} establish a $2$-MMS guarantee, which was subsequently improved to $4/3$ by Barman and Krishnamurthy~\cite{journals/teco/BarmanK20}, $11/9$ by Huang and Lu~\cite{conf/sigecom/HuangL21}, and $13/11$ by Huang and Segal-Halevi~\cite{conf/sigecom/HuangS23}.
The second is \emph{ordinal approximation}, which relaxes the benchmark by partitioning the chores into $d\le n$ bundles instead of $n$ bundles.
An allocation is $1$-out-of-$d$ MMS if every agent receives cost at most her MMS under this partition size.
Hosseini, Searns, and Segal-Halevi~\cite{conf/atal/HosseiniSS22} show that such allocations exist for $d=\lfloor3n/4\rfloor$, which was improved to $d=\lfloor9n/11\rfloor$ by Garg, Huang, and Segal-Halevi~\cite{conf/aaai/GargHS25}.
The third approach allows \emph{item disposal}: some chores may be left unallocated so that all agents receive cost at most their MMS, computed with respect to the original set of chores.
Barman, Rammohan, and Sethia~\cite{journals/corr/abs-2502-03789} show that discarding at most $2m/11+n$ chores suffices.
Recently, Ma and Wu~\cite{journals/corr/abs-2610-25320} improved this bound to $n-3$ for $n\ge4$.

Another prominent direction uses randomization to provide fairness guarantees both in expectation and in every realized allocation, commonly referred to as \emph{best-of-both-worlds} (BoBW) fairness.
An allocation lottery is ex-ante proportional if each agent's expected cost is at most $1/n$ of her total cost for all chores.
This guarantee alone is easy to achieve: we can simply assign all chores to a uniformly chosen agent.
However, the selected agent may then receive a bundle with cost far above her MMS.
It is therefore desirable to combine ex-ante proportionality with a strong ex-post fairness guarantee.
For chores, Babaioff, Ezra, and Feige~\cite{conf/wine/BabaioffEF22} observe that randomly permuting the agents in the picking sequence of Aziz, Li, and Wu~\cite{journals/mp/AzizLW24} yields ex-ante proportionality and ex-post $5/3$-MMS.
Feige and Huang~\cite{conf/sigecom/FeigeH23} improve the ex-post guarantee to $8/5$-MMS.

In this paper, we aim to achieve ex-ante proportionality while guaranteeing MMS for almost all agents in every outcome.
Since MMS cannot be guaranteed for everyone, we also impose a worst-case ex-post fairness guarantee for the remaining agents: namely, that removing at most one chore from any non-MMS agent's bundle makes her MMS.

\subsection{Our Results}

We show that ex-ante proportionality can be achieved together with MMS for all but a sublinear number of agents in every outcome.
For the remaining agents, we guarantee MMS up to one chore (MMS1), i.e., removing at most one assigned chore makes the agent's bundle MMS.
Using the $\tilde{O}(\cdot)$ notation to suppress polylogarithmic factors in $n$, we establish the following main result.

\begin{theorem}\label{thm:main}
Every instance of additive chores admits a finite ex-ante proportional lottery in which every allocation is MMS1 and all but $\tilde{O}(n^{3/4})$ agents are MMS.
\end{theorem}

The MMS1 guarantee allows us to remove at most one chore from each non-MMS agent's bundle to make every agent MMS.
Thus, every supported outcome becomes MMS after discarding at most $\tilde{O}(n^{3/4})$ chores.
Moreover, MMS1 immediately implies that every outcome is $2$-MMS.

We prove our result via a novel algorithmic framework that constructs a lottery over fractional allocations~\cite{conf/wine/BabaioffEF22,journals/corr/abs-2604-27211}.
Working with fractional allocations (as opposed to integral allocations directly) provides the flexibility needed to enforce strong structural properties before rounding.
We then apply unit-slot rounding~\cite{journals/mp/ShmoysT93,conf/wine/BabaioffEF22,journals/theoretics/FengL25,journals/corr/abs-2609-35280} to convert the fractional allocations into integral ones.

\smallskip

Our framework also yields the following guarantees by applying different structural constraints to the fractional allocations, where each row corresponds to an ex-ante proportional lottery.
\begin{center}
\small
\renewcommand{\arraystretch}{1.15}
\begin{tabular}{@{}p{.25\textwidth}p{.65\textwidth}@{}}
\toprule
Ex-post guarantee & Further guarantee\\
\midrule
$4/3$-MMS & MMS1 for every agent\\
$1$-out-of-$d$ MMS & $d=n-\tilde{O}\!\left(n^{2/3}\right)$\\
$\tilde{O}\!\left(n^{2/3}\right)$-AMMS & The non-MMS agent is chosen uniformly at random \\
\bottomrule
\end{tabular}
\end{center}

As a direct consequence, these lotteries improve the state of the art across several well-studied approximate MMS settings, including MMS with disposal, best-of-both-worlds fairness, MMS with subsidy, ordinal approximation, and all-but-one MMS (AMMS):
\begin{itemize}
    \item For MMS with disposal, we reduce the number of disposals from $n-3$~\cite{journals/corr/abs-2610-25320} to a sublinear function of $n$, while ensuring ex-ante proportionality before any chores are removed.
    \item For the best-of-both-worlds setting, our $4/3$-MMS lottery improves the ex-post $8/5$-MMS guarantee of Feige and Huang~\cite{conf/sigecom/FeigeH23} while preserving ex-ante proportionality. In addition, every outcome of our lottery is MMS1.
    \item Our main result yields a total MMS subsidy of $\tilde{O}(n^{3/4})$ under the standard normalization $c_i(e)\le1$, since each MMS1 agent can be made MMS with a subsidy of at most one dollar. This resolves an open problem posed in~\cite{journals/jair/LiuLSW24,conf/wine/WuZ24,conf/ifaamas/GargSW26,journals/corr/abs-2609-15208} and establishes a separation between MMS and PROP in total subsidy requirement~\cite{journals/opre/WuZZ26}.
    \item For ordinal approximation, we improve the $d=\lfloor9 n/11\rfloor$ guarantee for $1$-out-of-$d$ MMS allocations by Garg, Huang, and Segal-Halevi~\cite{conf/aaai/GargHS25} to $d=(1-o(1))n$ while simultaneously achieving ex-ante proportionality.
    \item An allocation is called $\alpha$-approximate all-but-one
    MMS ($\alpha$-AMMS) if $n-1$ agents receive bundles of cost at most their MMS, while the remaining agent ensures $\alpha$-MMS. 
    We improve upon the $\frac{(n+1)^2}{4n}$-AMMS bound of~\cite{conf/www/QiuWZZ26} by achieving $\tilde{O}(n^{2/3})$-AMMS while also guaranteeing ex-ante proportionality.
    Moreover, the exception agent is uniformly designated, so every fixed agent is MMS with probability at least $1-1/n$.
\end{itemize}

\subsection{A Technical Overview}

In what follows, we give an overview of the proof of our main result, assuming an identically ordered (IDO) instance and MMS value $\mu_i=1$ for every agent.

Many ex-ante proportional lotteries satisfy the \emph{equal-marginal property}, under which every chore is allocated to every agent with probability $1/n$.
Such a lottery can be obtained by assigning a $1/n$ fraction of every chore to each agent, organizing the fractions into unit-slots in decreasing cost order, and applying unit-slot rounding~\cite{journals/mp/ShmoysT93,journals/theoretics/FengL25,journals/corr/abs-2609-35280}.
However, standard unit-slot rounding may increase an agent's cost by up to one chore, which only guarantees MMS1.
To guarantee MMS for almost all agents, we need a stronger condition ensuring that the fractional bundles remain MMS even after rounding.
Since an agent receives at most one chore from each slot, it suffices to ensure that the sum of the maximum chore costs over all slots is at most one.
We call such fractional bundles (and the corresponding fractional allocations) \emph{rounding-resilient}.
We follow the paradigm of Babaioff, Ezra, and Feige~\cite{conf/wine/BabaioffEF22} and Babaioff, Feige, and Manaker Morag~\cite{journals/corr/abs-2604-27211}, which first constructs a lottery over fractional allocations and then applies unit-slot rounding to obtain integral allocations.
Our goal is to construct fractional allocations that are (nearly) rounding-resilient.

To illustrate how to construct such bundles, consider the idealized case where $q\approx \sqrt n$ is an integer dividing both $m$ and $n$.
We partition the ordered chores into blocks $B_1,\ldots,B_{m/q}$ of size $q$ and call the most costly chore $\hat e_j$ in each block $B_j$ its \emph{representative}.
Fix an agent $i$ and suppose that these representatives $\hat e_1,\ldots,\hat e_{m/q}$ can be packed into $n/q$ bins, each of total cost at most one.
For each bin $t$, we define a fractional bundle $x_{i,t}$ that assigns a $1/q$ fraction to every chore in a block whose representative belongs to bin $t$, and zero to every other chore.
For example, if bin $t$ contains $\hat e_3,\hat e_4$, and $\hat e_7$, then $x_{i,t}(e) = 1/q$ if $e \in B_3\cup B_4\cup B_7$, and $x_{i,t}(e) = 0$ otherwise.
Each such block forms a single unit-slot of total fraction $q\cdot(1/q)=1$, so the bin's cost bound ensures that $x_{i,t}$ is rounding-resilient.
By selecting each bundle with probability $q/n$, every chore receives expected fraction $(q/n)\cdot (1/q)=1/n$, ensuring ex-ante proportionality for agent $i$.
However, different agents may use different packings, and sampling their bundles independently may not yield a feasible fractional allocation.
If we could coordinate their lotteries into complete fractional allocations while preserving each agent's marginal distribution, rounding would immediately yield ex-ante proportionality and ex-post exact MMS.

Several obstacles arise in carrying out this plan.
First, the representatives may not fit into the required bins directly, but we show that they fit after setting aside blocks containing only $o(n)$ chores.
We choose $q$ to be a power of two so that we can build this packing through repeated halving steps.
We also rearrange the packing so that consecutive representatives are grouped into polynomially many segments, which bounds the number of distinct chore profiles we need to coordinate across agents.
Second, if $q$ does not divide $n$, we set aside another $o(n)$ chores from the last merged bin.
By an order-preserving shift, we can assume that all agents set aside the exact same set of $o(n)$ chores.
We then restore these excluded chores back into the fractional bundles while keeping each agent's expected share of every chore equal to $1/n$ and the cost of every fractional bundle at most one.
Third, to coordinate the agents' choices and allocate all chores completely, we reserve $o(n)$ agents at the beginning.
Their shares leave room to absorb small concentration deviations in the other agents' allocations, and they receive the leftover fractions (including fractions of the chores set aside).
Under our coordinated sampling, almost every non-reserved agent receives a single prepared fractional bundle (which is almost rounding-resilient), while only $o(n)$ agents receive a convex mixture of multiple bundles.
After rounding, any non-reserved agent who receives a single prepared bundle is guaranteed to be MMS unless she receives one of the chores set aside.
Consequently, only the reserved agents, the agents receiving mixed bundles, and the recipients of the chores set aside can fail to be MMS.
Together, these account for only $o(n)$ agents, while every agent remains MMS1.

\subsection{Other Related Work}

Given the vast literature on fair division, we focus here on the most closely related work and refer the reader to recent surveys~\cite{journals/sigecom/AzizLMW22,journals/ai/AmanatidisABFLMVW23} for a comprehensive overview.

\paragraph{MMS approximation for goods.}
For goods, an $\alpha$-MMS allocation requires that every agent receives value at least $\alpha$ times her MMS, where $\alpha\le1$.
Kurokawa, Procaccia, and Wang~\cite{journals/jacm/KurokawaPW18} show that MMS allocations are not guaranteed to exist, and $2/3$-MMS allocations always exist for additive valuations.
This approximation ratio was improved to $3/4$ by \cite{journals/moor/GhodsiHSSY21} and Garg and Taki~\cite{journals/ai/GargT21}.
Akrami and Garg~\cite{conf/soda/AkramiG24} obtain the first constant improvement beyond $3/4$, and \cite{conf/soda/HeidariKSS26} further improve the ratio to $10/13$, and Huang and Zhou~\cite{journals/corr/abs-2511-13056} establish the existence of $7/9$-MMS allocations.
For ordinal approximation, Hosseini, Searns, and Segal-Halevi~\cite{journals/jair/HosseiniSS22} show that $1$-out-of-$\lfloor3n/2\rfloor$ MMS allocations always exist.
Akrami, Garg, Sharma, and Taki~\cite{conf/sigecom/AkramiGST24} improve this bound to $1$-out-of-$4\lceil n/3\rceil$ MMS, and Ezra and Garbuz~\cite{journals/corr/abs-2609-33234} further obtain $1$-out-of-$(1+o(1))n$ MMS allocations.
On the negative side, Ezra and Garbuz~\cite{journals/corr/abs-2609-15085} show that, for both goods and chores, the multiplicative approximation guarantee cannot approach MMS at an inverse-polynomial rate in $n$.

\paragraph{Further BoBW guarantees.}
For goods, \cite{journals/ior/AzizFSV24} combine ex-ante envy-freeness with ex-post envy-freeness up to one item, while Babaioff, Ezra, and Feige~\cite{conf/wine/BabaioffEF22} guarantee ex-ante proportionality, ex-post PROP1, and $1/2$-MMS.
Ezra and Garbuz~\cite{journals/corr/abs-2609-33234} preserve ex-ante proportionality while guaranteeing all but $O(n^{2/3}(\log n)^{1/3})$ agents MMS or $1$-out-of-$(n+O(n^{2/3}(\log n)^{1/3}))$ MMS in every outcome.
Their weighted analogs preserve ex-ante weighted proportionality and use AnyPrice Share (APS) for sufficiently small entitlements.
In contrast, our lottery bounds the number of MMS exceptions in every realized outcome and guarantees MMS1 for everyone, which yields a sublinear disposal guarantee.
Prakash H.V. and Nimbhorkar~\cite{conf/atal/HVN24} obtain ex-ante weighted stochastic-dominance envy-freeness and ex-post weighted stochastic-dominance proportionality up to one item for both goods and chores.

\section{Preliminaries}

For a positive integer $t$, let $[t]=\{1,\ldots,t\}$.
We consider a set $M=\{e_1,\ldots,e_m\}$ of indivisible chores and a set $N=[n]$ of agents, where each agent $i$ has a nonnegative additive cost function $c_i$.
An allocation $\bX=(X_1,\ldots,X_n)$ is a partition of $M$ into $n$ bundles, where agent $i$ receives bundle $X_i$.
A fractional allocation is a tuple $x=(x_1,\ldots,x_n)$ of fractional bundles $x_i$, where $x_i(e)\in[0,1]$ is the fraction of chore $e$ assigned to agent $i$, subject to $\sum_{i\in N} x_i(e)\le 1$ for every $e\in M$ (with equality for a complete fractional allocation).
The cost of a fractional bundle $x_j$ to agent $i$ is $c_i(x_j)=\sum_{e\in M}c_i(e)\cdot x_j(e)$.
A lottery is a probability distribution over integral allocations: an ex-ante guarantee bounds each agent's expected cost under this distribution, while an ex-post guarantee holds for every allocation in the support.

For every positive integer $d$, let $\Pi(M,d)$ be the collection of all partitions $\mathbf{P}=(P_1,\ldots,P_d)$ of $M$ into $d$ bundles.
The $1$-out-of-$d$ maximin share for agent $i$ is defined as 
\begin{equation*}
 \operatorname{MMS}_i^d(M)
 =\min_{\mathbf{P} \in \Pi(M,d)}
   \left\{\max_{j\in[d]}\left\{c_i\!\left(P_j\right)\right\}\right\},
 \qquad
 \mu_i=\operatorname{MMS}_i^n(M).
\end{equation*}

An allocation $\bX$ is called MMS if $c_i(X_i)\le\mu_i$ for every agent $i \in N$.
Let $\PROP_i=c_i(M)/n$ denote agent $i$'s proportional share, and $\kappa_i=\max_{e\in M}\{c_i(e)\}$ her maximum single-chore cost.
Without loss of generality (w.l.o.g.), $m = |M|$ may be assumed sufficiently large by adding dummy chores with zero cost.
If $\mu_i=0$ for some agent $i$, assigning all chores to that agent results in zero cost for every agent. 
We henceforth assume $\mu_i>0$ for all $i\in N$.
By adding infinitesimal noises to the costs, we can further assume that $c_i(e)$ are different and positive for all agent $i\in N$ and chore $e\in M$.
By scaling each cost function, we assume $\mu_i=1$, which implies $\PROP_i\le 1$ and $\kappa_i\le 1$.
A lottery is ex-ante proportional if $\mathbb E[c_i(X_i)]\le\PROP_i$ for every agent $i$.
An allocation is $\alpha$-MMS if $c_i(X_i)\le\alpha$ for every agent, $\alpha$-AMMS if it is $\alpha$-MMS and gives MMS to all agents except possibly one, and $1$-out-of-$d$ MMS if $c_i(X_i)\le\operatorname{MMS}_i^d(M)$ for every agent.

\begin{definition}[MMS1]
A bundle $X$ is \emph{MMS up to one chore (MMS1)} for agent $i$ if either $X=\varnothing$, or there exists a chore $e\in X$ such that $c_i\!\left(X\setminus \{e\}\right)\le 1$.
An allocation is MMS1 if every agent receives an MMS1 bundle.
\end{definition}

Given any MMS1 allocation, deleting at most one chore from each non-MMS agent's bundle makes the allocation MMS.
Therefore, an upper bound on the number of non-MMS agents immediately gives an upper bound on the number of chore disposals needed to recover MMS.

We call an instance identically ordered (IDO) if we can index its chores so that
\begin{equation*}
 c_i\!\left(e_1\right)\ge c_i\!\left(e_2\right)\ge\cdots\ge c_i\!\left(e_m\right)\qquad \forall i\in N.
\end{equation*}
We use the standard IDO reduction; see, e.g.,~\cite{conf/aaai/GargHS25}.

\begin{lemma}[IDO reduction]\label{lem:ido-reduction}
Given any instance $\left(N,M,c\right)$, we can create another IDO instance in which $c_i(e_t)$ equals the $t$-th highest cost of agent $i$ in the original instance.
Any allocation for the IDO instance can be converted into an allocation for the original instance, in which the cost of every agent is at most that in the IDO instance.
Each assigned rank is replaced by a distinct actual chore of no greater cost to its owner.
\end{lemma}
\begin{proof}
Process assigned ranks from $m$ down to $1$.
At rank $t$, its owner chooses the cheapest unassigned actual chore.
Since at least $t$ chores remain, and at most $t-1$ cost her more than her $t$-th highest cost,
her choice costs at most $c_i(e_t)$ in the IDO instance.
The itemwise comparison preserves MMS1 and does not increase the number of non-MMS agents.
\end{proof}

Henceforth we work only on IDO instances.
We write $e_j$ for the $j$-th most costly chore.

\section{Lottery on Allocations with Sublinear Disposal}\label{sec:sublinear-disposal}

Since we aim for the sublinear guarantee, we may assume without loss of generality that $n$ is sufficiently large.
For any bounded $n$, $\tilde{O}(n^{3/4})$ is equivalent to $O(n)$, and ex-ante PROP and ex-post MMS1 is implied by ex-ante PROP and ex-post PROP1~\cite[Theorem~5.3]{conf/atal/HVN24}.

Fix a sufficiently large absolute constant $C_1$ and set
\begin{equation*}
 q=2^{\lfloor\log_2\sqrt{n\log n}\rfloor},
 \qquad
 k=\lceil C_1 n^{3/4}(\log n)^{1/4}\rceil.
\end{equation*}
By definition, $q$ is a power of two, with $q=\tilde O(n^{1/2})$ and $k=\tilde O(n^{3/4})$.
For sufficiently large $n$, both parameters lie in $\left[1,n\right)$.
Fix any $k$-element subset $N_r\subseteq N$ and let $N_o=N\setminus N_r$.
The agent set $N$ is thus partitioned into the \emph{reserved} agents $N_r$ and the \emph{ordinary} agents $N_o$.

Since the choice of $N_r$ is arbitrary, our construction holds for every fixed $k$-subset $N_r\subseteq N$. 
Consequently, $N_r$ may also be drawn randomly before applying the construction conditioned on that choice.
Intuitively, we aim to ensure that the ordinary agents in $N_o$ satisfy the MMS bound after rounding, while the reserved agents in $N_r$ absorb the remaining chore fractions.
However, the reserved agents are not the only possible non-MMS agents: a small number of ordinary agents may also violate the MMS bound, and our analysis accounts for these exceptions as well.
Instead of constructing a lottery over integral allocations directly, we begin with a lottery over fractional allocations and subsequently round them into integral ones.
Specifically, we construct weighted fractional bundles for each ordinary agent, coordinate their weights while assigning the complementary fractions to the reserved agents, and finally round all bundles jointly.
On IDO instances, the resulting lottery assigns every chore to every agent with probability $1/n$.
The IDO reduction preserves ex-ante proportionality, although it may change the marginal probabilities of individual chores.

Our construction of the lottery proceeds in three main steps:
\begin{itemize}
    \item First, for each ordinary agent $i\in N_o$, we construct a lottery over fractional bundles $x_{i,1},x_{i,2},\ldots,x_{i,L}$.
    These bundles are designed to be rounding-resilient after excluding $o(n)$ chores, which means that unit-slot rounding yields an integral bundle of cost at most $1$ for agent $i$ (i.e., satisfying MMS).
    \item Since the bundle lotteries for different ordinary agents are constructed independently, we next coordinate their sampling probabilities to form a feasible joint fractional allocation while preserving each ordinary agent’s expected fraction of every chore, and then divide the remaining unallocated fraction of each chore evenly among the reserved agents in $N_r$ to make the allocation complete.
    \item Finally, the fractional allocations are rounded into integral ones by grouping the fractions into unit-slots and applying the Birkhoff--von Neumann theorem.
    Because our construction ensures $c_i(x_i)\leq 1$ for every agent $i$'s assigned fractional bundle $x_i$, every agent satisfies MMS1 after rounding. 
    Moreover, by the rounding-resilience property, almost all ordinary agents satisfy MMS after rounding.
\end{itemize}

\subsection{Lottery of Fractional Bundles}
\label{sec:fractional-allocations}
\begin{figure}[htbp]
\centering
\begin{tikzpicture}[x=1cm,y=1cm,font=\small,
 block/.style={draw=restoreblue!65,rounded corners=2pt,line width=.65pt},
 rep/.style={draw=restoreblue,fill=restoreblue!14,rounded corners=2pt,
   minimum width=.67cm,minimum height=.47cm,inner sep=2pt},
 bin/.style={draw=restoreblue!75,rounded corners=3pt,line width=.7pt,
   minimum width=2.95cm,minimum height=1.65cm,inner sep=3pt},
 binrep/.style={rep,minimum width=1.04cm,minimum height=.43cm,
   inner sep=1pt,font=\footnotesize}]
 \foreach \x/\lab/\first/\last in {
   1.50/{B_1}/{\hat e_1}/{e_q},
   4.90/{B_2}/{\hat e_2}/{e_{2 q}},
   8.30/{B_j}/{\hat e_j}/{e_{j\cdot q}},
   12.65/{B_{m/q}}/{\hat e_{m/q}}/{e_m}} {
   \draw[block] (\x-1.15,3.05) rectangle (\x+1.15,3.76);
   \node at (\x,4.03) {$\lab$};
   \node[rep] at (\x-.65,3.405) {$\first$};
   \node at (\x+.20,3.405) {$\cdots$};
   \node at (\x+.80,3.405) {$\last$};
 }
 \node[rep,draw=gray!65,fill=gray!14,text=gray!85]
   at (.85,3.405) {$\hat e_1$};
 \node[text=gray!80,font=\footnotesize] at (1.50,2.75) {excluded};
 \node at (10.475,3.405) {$\cdots$};
 \draw[restoreblue!60] (3.75,2.86)--(3.75,2.75)--(6.05,2.75)--(6.05,2.86);
 \node[font=\footnotesize,text=restoreblue!90!black] at (4.90,2.53) {$q$ chores};
 \draw[->,restoreblue,line width=.85pt] (7.075,2.60)--(7.075,1.40);
 \node[anchor=east,font=\footnotesize] at (6.825,1.73) {Lemma~\ref{lem:core-reduction}};
 \node[anchor=west,font=\footnotesize,text=gray!85] at (7.325,1.73)
   {$O(\sqrt L)$ excluded};
 \node[bin] (binone) at (1.50,0) {};
 \node[bin] (bintwo) at (4.90,0) {};
 \node[bin] (binlast) at (9.25,0) {};
 \node[bin,draw=restoreorange] (binlasttwo) at (12.65,0) {};
 \node at (1.50,1.10) {$1$};
 \node at (4.90,1.10) {$2$};
 \node at (9.25,1.10) {$L$};
 \node at (12.65,1.10) {$L+1$};
 \node at (7.075,0) {$\cdots$};
 \foreach \x/\first/\second/\third in {
   1.50/{\hat e_2}/{\hat e_3}/{\hat e_4},
   4.90/{\hat e_5}/{\hat e_6}/{\hat e_7},
   9.25/{\hat e_j}/{\hat e_{j+1}}/{\hat e_{j+2}},
   12.65/{\hat e_{m/q-2}}/{\hat e_{m/q-1}}/{\hat e_{m/q}}} {
   \node[binrep] at (\x-.72,.35) {$\first$};
   \node[binrep] at (\x+.72,.35) {$\second$};
   \node[binrep] at (\x-.72,-.35) {$\third$};
   \node at (\x+.72,-.35) {$\cdots$};
 }
 \node[font=\footnotesize] at (1.50,-1.10) {$\le1$};
 \node[font=\footnotesize] at (4.90,-1.10) {$\le1$};
 \node[font=\footnotesize] at (9.25,-1.10) {$\le1$};
 \node[font=\footnotesize,text=restoreorange!90!black] at (12.65,-1.10) {$\le h/q$};
 \draw[gray!65] (.025,-1.40)--(.025,-1.52)--(10.725,-1.52)--(10.725,-1.40);
 \node[font=\footnotesize] at (5.375,-1.80) {$L\text{ bins of cost at most }1$};
\end{tikzpicture}
\caption{Blocks, representative chores, and a schematic partition from Lemma~\ref{lem:core-reduction} when $h>0$.}
\label{fig:blocks-and-bins}
\end{figure}
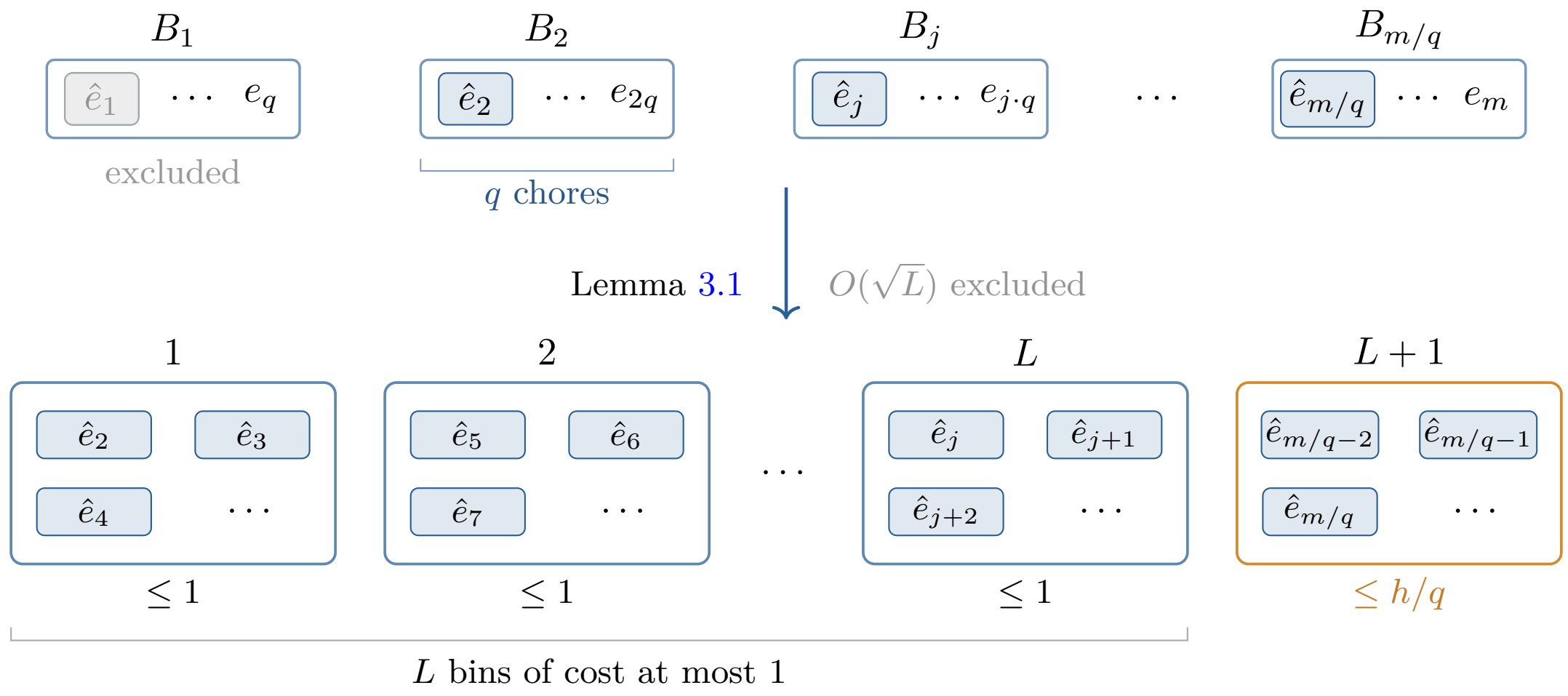

In this subsection, we construct a lottery over fractional bundles for a fixed ordinary agent $i$.
By adding zero-cost dummy chores if necessary, we may assume that $m$ is a multiple of $q$. 
Partition $M$ into $m/q$ blocks of size $q$.
Recall that we focus on IDO instances.
For each $j\in[m/q]$, we define the block $B_j$ and its \emph{representative chore} $\hat e_j$ by
\begin{equation*}
 B_j=\{e_{(j-1) q+1},\ldots,e_{j q}\}
 \quad \text{and} \quad
 \hat e_j=e_{(j-1) q+1},
\end{equation*}
so that every chore in $B_j$ costs at most $c_i(\hat e_j)$.
We further set
\begin{equation*}
 L=\lfloor n/q\rfloor \quad \text{and} \quad h=n-q\cdot L \quad (0\le h<q),
\end{equation*}
where $L=\tilde O(\sqrt{n})$ under our choice of $q$.
Here, $L$ denotes the number of unit-capacity bins (i.e., each of total cost at most $1$ under the agent's cost function) used to pack the representative chores.
When $h>0$, we allow one additional bin of capacity $h/q$, which is later merged into bin $L$.
The resulting $L$ bins are then used to construct the agent's $L$ candidate fractional bundles (see Figure~\ref{fig:blocks-and-bins} for an illustration of the blocks and output bins).
We leave $B_1$ unassigned at this stage and partition the remaining representative chores using the following lemma.

\begin{lemma}\label{lem:core-reduction}
For each agent $i\in N$, all but $O(\sqrt L)$ of the representative chores $\hat e_2,\ldots,\hat e_{m/q}$ can be partitioned into $L$ bins of cost at most $1$ each with respect to $c_i$, plus one additional bin of cost at most $h/q$ if $h>0$.\footnote{Note that both the excluded representative chores and the partition may depend on agent $i$.}
\end{lemma}
\begin{proof}[Proof Sketch]
We start with an MMS partition and replace chores by cheaper representative copies, obtaining a feasible packing of $q$ copies of every representative except $\hat e_1$ into $n$ unit-capacity bins.
We then repeatedly retain half the copies of each representative and merge bins in pairs.
We coordinate these choices so that excess load in one bin is offset by spare capacity in another.
This balancing limits the number of bins that need exclusions.
Only a few copies must be temporarily set aside at each step, and subsequent halving also reduces their number.
Finally, excluding these copies leaves one copy of each remaining representative packed into the bins with the required capacities.
See Appendix~\ref{app:core-reduction} for the full proof.
\end{proof}

For each ordinary agent $i\in N_o$, we take the partition given by Lemma~\ref{lem:core-reduction}. 
If $h>0$, we merge the additional bin into bin $L$, which yields $L$ bins with cost bounds $1,\ldots,1,1+h/q$.
Since these bins contain only the representative chores, we next replace each non-excluded representative chore $\hat e_j$ with all chores in its original block $B_j$, and let $G_{i,t}$ denote the union of the blocks placed in the resulting bin $t\in[L]$.
By construction, the sets $G_{i,1},\ldots,G_{i,L}$ are pairwise disjoint and satisfy $c_i(G_{i,t})\le q$ for all $t<L$ and $c_i(G_{i,L})\le q+h$, as each block contains $q$ chores.

Next, for a fixed ordinary agent $i\in N_o$, we use the sets
$G_{i,1},\ldots,G_{i,L}$ to construct fractional bundles $y_{i,t}$ and specify their coefficients $\eta_{i,t}$ for $t\in[L]$, which ensures that each bundle has cost at most $1$.
Specifically, for each $t\in[L]$, we initially set
\begin{equation*}
 \left\{
 \begin{aligned}
  \eta_{i,t} &= \frac{q}{n},   &\quad y_{i,t}(e) &= \frac{1}{q}\cdot\mathbf{1}(e\in G_{i,t})   &&\quad \text{if } t<L, \\[4pt]
  \eta_{i,t} &= \frac{q+h}{n}, &\quad y_{i,t}(e) &= \frac{1}{q+h}\cdot\mathbf{1}(e\in G_{i,t}) &&\quad \text{if } t=L.
 \end{aligned}
 \right.
\end{equation*}

These fractional bundles are called \emph{base bundles}, as they serve as the starting point before shifting and restoring the excluded chores.
Clearly, $c_i(y_{i,t}) \leq 1$ holds for every $t \in [L]$.
Moreover, since each unmerged set $G_{i,t}$ (for $t<L$) is a union of blocks whose representative chores have total cost at most $1$, receiving at most one chore per block after rounding ensures that agent $i$'s rounded cost remains at most $1$.
This rounding-resilience property is crucial for bounding both the number of non-MMS agents and the number of chore disposals in our subsequent analysis.

For the merged bundle $y_{i,L}$, we exclude the $\min\{q+h,\vert{}G_{i,L}\vert{}\}$ highest-cost chores in $G_{i,L}$ (breaking ties by chore index) by setting their fractions in $y_{i,L}$ to zero, while keeping $\eta_{i,L}$ unchanged even if the bundle becomes empty.
Note that these additional exclusions ensure that the base bundle has cost at most $1$ after rounding, as we will show in Section~\ref{sec:rounding}.

Let $Z_i$ be the union of $B_1$ (containing $q$ chores), all blocks whose representatives were excluded in Lemma~\ref{lem:core-reduction} (contributing $O(q\cdot \sqrt{L})$ chores), and the chores excluded from $G_{i,L}$ (fewer than $2q$ chores).
Thus, $Z_i$ is nonempty and $\vert{}Z_i\vert{}=O(q\cdot \sqrt{L})$, and every chore in $M\setminus Z_i$ belongs to the support of exactly one base bundle.
Moreover, the coefficients satisfy $\sum_{t=1}^L\eta_{i,t}=1$ and $\eta_{i,t}\in [q/n, 2q/n]$.
Since multiplying each surviving chore's fraction by its bundle's coefficient yields $1/n$, it follows that
\begin{equation*}
 \sum_{t=1}^L\eta_{i,t}\cdot y_{i,t}(e)=\frac{1-\mathbf{1}(e\in Z_i)}{n}
 \quad\text{and}\quad
 0\le y_{i,t}(e)\le\frac{1}{q}.
\end{equation*}

\begin{figure}[htbp]
\centering
\begin{tikzpicture}[x=.76cm,y=.70cm,font=\small]
  \node[anchor=east,align=right] at (.30,2.05) {Before};
  \node[anchor=east,align=right] at (.30,0) {After};
  \foreach \u in {1,...,12} {
    \node[draw=restoreblue,fill=restoreblue!8,minimum width=.64cm,
      minimum height=.48cm,inner sep=0pt] at (\u,2.05) {$e_{\u}$};
    \node[draw=restoreblue,fill=restoreblue!8,minimum width=.64cm,
      minimum height=.48cm,inner sep=0pt] at (\u,0) {$e_{\u}$};
  }
  \foreach \u in {1,2,5,6} {
    \node[draw=gray!65,fill=gray!16,text=gray!85,minimum width=.64cm,
      minimum height=.48cm,inner sep=0pt] at (\u,2.05) {$e_{\u}$};
  }
  \node[draw=gray!65,dashed,fill=gray!8,minimum width=.64cm,
    minimum height=.48cm,inner sep=0pt] at (12,2.05) {$e_{12}$};
  \node[font=\scriptsize,text=gray!80] at (12,2.63) {removed};
  \foreach \u in {1,...,5} {
    \node[draw=restoreorange,fill=restoreorange!18,minimum width=.64cm,
      minimum height=.48cm,inner sep=0pt] at (\u,0) {$e_{\u}$};
  }
  \foreach \old/\new in {3/6,4/7,7/8,8/9,9/10,10/11,11/12} {
    \draw[->,restoreblue!75,line width=.55pt]
      (\old,1.68)--(\new,.37);
  }
  \draw[restoreorange,line width=.7pt] (.58,-.52)--(.58,-.68)--(5.42,-.68)--(5.42,-.52);
  \node[anchor=north,text=restoreorange!90!black] at (3,-.78) {$H=\left\{e_1,\ldots,e_5\right\}$};
  \node[anchor=north] at (8,-.65) {$c_i\!\left(e_{K+j}\right)\le c_i\!\left(e_{r_j}\right)$};
\end{tikzpicture}
\caption{For $m=12$ and $K=5$, the excluded set is $Z_i=\{e_1,e_2,e_5,e_6\}$, which contains the first block $B_1=\{e_1,e_2\}$. The original fraction on $e_{12}$ is removed.}
\label{fig:common-set-restoration}
\end{figure}
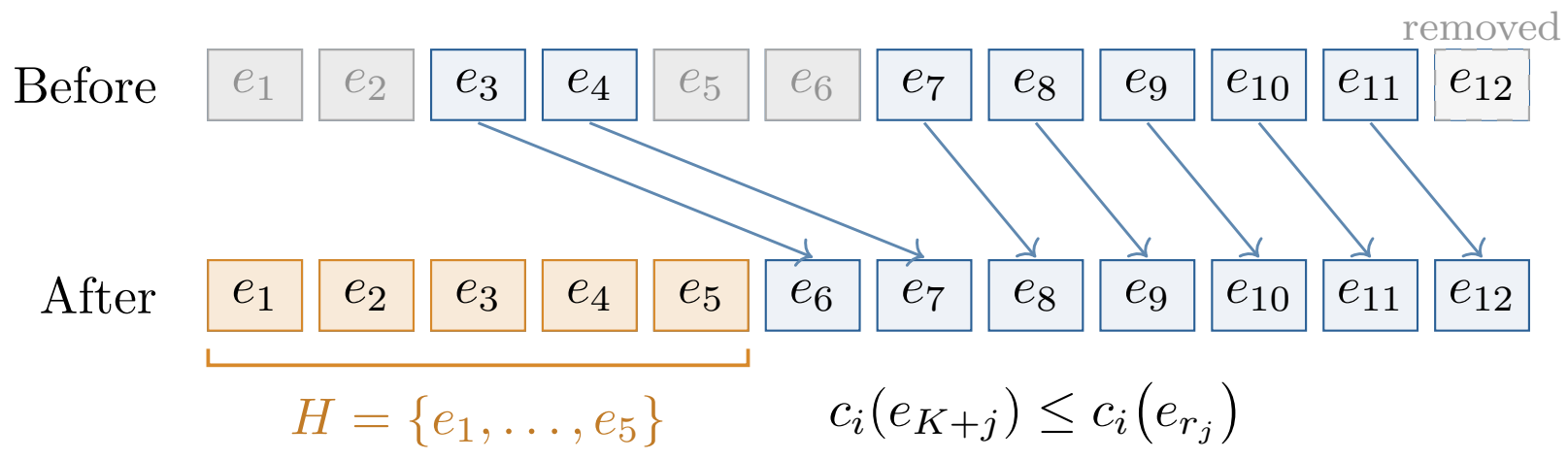

\paragraph{A Common Set of Excluded Chores $H$.}
In the construction above, the excluded sets $Z_i$ may vary across different ordinary agents, but we can align them to a common set $H$ without increasing the cost of any bundle.
Specifically, let $K=\max_{i\in N_o} \{|Z_i|\}$ and let $H=\{e_1,\ldots,e_K\}$ be the $K$ highest-cost chores, where $K=O(q\cdot \sqrt{L})=\tilde O(n^{3/4})$ for our choices of $k$ and $q$.
For each ordinary agent $i\in N_o$, we list the surviving chores $M\setminus Z_i$ in increasing index order as $e_{r_1},e_{r_2},\ldots$ and shift each chore $e_{r_j}$ to $e_{K+j}$, removing it from all fractional bundles of agent $i$ whenever $K+j$ exceeds $m$.
Since $|Z_i|\le K$, there are at least $m-K$ surviving chores, and the inequality $r_j\le K+j$ ensures $c_i(e_{K+j})\le c_i(e_{r_j})$ for all $j\in[m-K]$.
The coefficients $\eta_{i,t}$ remain unchanged, and we continue to write $y_{i,t}$ for the shifted fractional bundles.
The resulting fractional bundles yield an expected fraction of $1/n$ on $M\setminus H$ and zero on $H$, with each bundle assigning at most a $1/q$ fraction of every chore.
Figure~\ref{fig:common-set-restoration} illustrates this shifting operation.

We next determine the fractions assigned to the chores in $H$.
A natural attempt is to divide these chores evenly among the reserved agents alone, but doing so could push their fractional costs above $1$ when $c_i(y_{i,t})$ is substantially smaller than $1$ for some ordinary agents.
To prevent this, we also allocate fractions of $H$ to each ordinary agent $i\in N_o$ by augmenting her base bundles $y_{i,t}$ in proportion to their slacks $1-c_i(y_{i,t})$, ensuring that every completed bundle still has cost at most $1$ and every chore in $H$ receives an expected fraction of $1/n$.
Since $H$ contains the highest-cost chores, $c_i(H)$ is strictly positive, and thus we have 
\begin{equation*}
 \sum_{t=1}^L\eta_{i,t} \cdot \bigl(1-c_i(y_{i,t})\bigr)
 =1-\frac{c_i(M\setminus H)}{n}
 =1-\PROP_i+\frac{c_i(H)}{n}\ge\frac{c_i(H)}{n}>0.
\end{equation*}
Accordingly, for each $t\in[L]$, we define the completed bundle $x_{i,t}$ by setting
\begin{align*}
 \tau_{i,t}
 &=\frac{1-c_i(y_{i,t})}{\sum_{j=1}^L\eta_{i,j}\cdot\bigl(1-c_i(y_{i,j})\bigr)}\cdot\frac1n
 =\frac{1-c_i(y_{i,t})}{n(1-\PROP_i)+c_i(H)},
 \ \ x_{i,t}(e)=
 \begin{cases}
  y_{i,t}(e) & \text{if } e\notin H, \\[2pt]
  \tau_{i,t} & \text{if } e\in H.
 \end{cases}
\end{align*}
Note that the expected assigned fraction of each chore in $H$ is $\sum_{t=1}^L\eta_{i,t}\cdot\tau_{i,t} = 1/n$.

Since $\PROP_i\le1$, we also obtain
\begin{align*}
 c_i(x_{i,t})
&=c_i(y_{i,t})+c_i(H)\cdot\tau_{i,t}=c_i(y_{i,t})+\frac{c_i(H)\cdot\left(1-c_i(y_{i,t})\right)}{n\cdot\left(1-\PROP_i\right)+c_i(H)}
\\&\le c_i(y_{i,t})+\left(1-c_i(y_{i,t})\right)=1.
\end{align*}

Recall that before assigning fractions of $H$, any base bundle $y_{i,t}$ with $t<L$ guarantees a rounded cost of at most $1$ whenever agent $i$ receives at most one chore from each block.
However, adding positive fractions of the chores in $H$ may break this rounding-resilience property, namely when an agent actually receives a chore from $H$ after rounding.
The good news is, under our parameter choices, $H$ has sublinear size, which yields sublinear bounds on both the number of resulting non-MMS agents and the number of chore disposals they require.

We further have $0\le x_{i,t}(e)\le1/q$ for every chore $e\in M$ because
\begin{equation*}
 \tau_{i,t} = \frac{1-c_i(y_{i,t})}{n\cdot\left(1-\PROP_i\right)+c_i(H)}
 \le\frac1{n\cdot\eta_{i,t}}\le\frac1q.
\end{equation*}

\paragraph{Summary.}
Step~1 produces a common set $H=\{e_1,\ldots,e_K\}$ of size $K=O(q\cdot \sqrt{L})$ and, for each agent $i\in N_o$, the fractional bundles $x_{i,1},\ldots,x_{i,L}$ with coefficients $\eta_{i,1},\ldots,\eta_{i,L}$ satisfy
\begin{equation*}
 \sum_{t=1}^L\eta_{i,t}=1,\qquad
 \frac{q}{n}\le\eta_{i,t}\le\frac{2q}{n},\qquad
 \sum_{t=1}^L\eta_{i,t}\cdot x_{i,t}(e)=\frac{1}{n},\qquad
 c_i(x_{i,t})\le 1,\qquad
 0\le x_{i,t}(e)\le\frac{1}{q}
\end{equation*}
for $t\in[L]$ and $e\in M$.
Note that this step specifies only a marginal distribution for each ordinary agent and leaves the reserved agents unassigned, so these individual lotteries do not yet form a valid joint allocation across all agents.
In the next step, we coordinate these marginal distributions to form a joint lottery over complete fractional allocations.

\subsection{Coordinating the Lotteries}
\label{sec:selection}
Although Step~1 provides a lottery over fractional bundles for each ordinary agent, sampling these bundles independently could assign excessive fractions of some chores while assigning too little of others.
We therefore carefully coordinate the bundle coefficients and allocate the remaining chore fractions to the reserved agents.
Recall that each ordinary agent receives an expected fraction of $1/n$ for every chore.
With $n-k$ ordinary agents in total, the expected total fraction of each chore $e$ assigned to $N_o$ is $1-k/n$.
Writing $\eta=(\eta_{i,t})_{i\in N_o,t\in[L]}$ for the vector of coefficients, we replace $\eta$ with a random vector $\widehat{\eta}\geq 0$ satisfying $\sum_{t=1}^L\widehat{\eta}_{i,t}=1$ for all $i\in N_o$.
Under $\widehat{\eta}$, the total fraction of chore $e$ assigned to the ordinary agents is
\begin{equation*}
 x(e):=\sum_{i\in N_o}\sum_{t=1}^L\widehat{\eta}_{i,t}\cdot x_{i,t}(e).
\end{equation*}
Note that $x(e)$ is a random variable depending on $\widehat{\eta}$, and need not equal its expectation $1-k/n$.
Our goal is to ensure $\mathbb E\left[\widehat{\eta}_{i,t}\right]=\eta_{i,t}$ for all $i$ and $t$, $x(e)\le1$ for every chore $e\in M$, and the $k$ cost equations for the reserved agents:
\begin{equation*}
 \sum_{i\in N_o,t}\widehat{\eta}_{i,t}\cdot c_j(x_{i,t})
 =\left(n-k\right)\cdot\PROP_j\qquad\forall j\in N_r.
\end{equation*}
Note that the equations ensure that the complementary fractions left for the reserved agents have total cost at most $k\cdot\PROP_j$ for every $j\in N_r$.

While setting $\widehat{\eta}=\eta$ trivially satisfies all of these properties, $\eta$ has entirely fractional coordinates by construction.
Intuitively, fractional coefficients mix several bundles, making it harder to guarantee MMS after rounding.
We thus introduce the random vector $\widehat{\eta}$ so that for almost all ordinary agents, $\widehat{\eta}_{i,t}$ takes binary values in $\{0,1\}$, making the choice of fractional bundle deterministic under any realization of $\widehat{\eta}$.
This property is crucial because if $\widehat{\eta}_{i,t}=1$, agent $i$ receives a single prepared fractional bundle $x_{i,t}$. 
Our rounding procedure (introduced in the next step) then gives her an MMS bundle whenever she receives no chore from $H$.

This motivates partitioning the ordinary agents according to whether their realized coefficients select a single bundle.
Specifically, under a fixed realization of $\widehat{\eta}$, we call an ordinary agent $i\in N_o$ \emph{pure} if exactly one coefficient $\widehat{\eta}_{i,t}$ is nonzero (and hence equal to $1$), and \emph{mixed} otherwise.
Assuming $q\ge 4\log n$ and $k\ge C_2\cdot n\sqrt{(\log n)/q}$ for a sufficiently large constant $C_2$, the following lemma provides the desired random vector $\widehat{\eta}$.
The proof is deferred to Appendix~\ref{app:joint-selection}.

\begin{lemma}
\label{lem:reserve-coordination}
There exists a finite distribution over coefficient vectors $\widehat{\eta}$ satisfying the conditions above, such that every outcome in its support yields at most $2(k+1)$ mixed agents.
\end{lemma}

Intuitively, Lemma~\ref{lem:reserve-coordination} consolidates the randomness of the individual lotteries into a single correlated vector $\widehat{\eta}$ that coordinates the sampling of fractional bundles while preserving all required marginal and cost invariants.
We sample $\widehat{\eta}$ from this distribution and set
\begin{equation*}
 x_i(e)=\sum_{t=1}^L\widehat{\eta}_{i,t}\cdot x_{i,t}(e)
 \qquad, \forall i\in N_o.
\end{equation*}
Since $x(e) = \sum_{i\in N_o} x_i(e) \leq 1$ for every chore $e\in M$, we can allocate an equal share of the remaining unassigned fraction to each reserved agent:
\begin{equation*}
 x_j(e)=\frac{1-x(e)}{k}
 \qquad, \forall j\in N_r,
\end{equation*}
which yields a complete fractional allocation $x=(x_1,\ldots,x_n)$.

For every ordinary agent $i\in N_o$, the bundle cost bounds imply
\begin{equation*}
 c_i(x_i)=\sum_{t=1}^L\widehat{\eta}_{i,t}\cdot c_i(x_{i,t})
 \le\sum_{t=1}^L\widehat{\eta}_{i,t}=1.
\end{equation*}
Similarly, for every reserved agent $j\in N_r$, the cost equations ensure
\begin{equation*}
 c_j(x_j)
 =\frac{c_j(M)-\sum_{i\in N_o}c_j(x_i)}{k}
 =\frac{c_j(M)-(n-k)\cdot\PROP_j}{k}
 =\PROP_j\le 1.
\end{equation*}
Thus, every agent's fractional cost is at most $1$.

\paragraph{Summary.}
The coordination step produces a lottery over complete fractional allocations that preserves each ordinary agent’s expected fraction of every chore.
This joint sampling is governed by the random vector $\widehat{\eta}$, whose coordinates are binary for all but at most $2(k+1)$ ordinary agents.
Once $\widehat{\eta}$ is realized, the complete fractional allocation $x=(x_1,\ldots,x_n)$ is determined, with each agent $i\in N$ receiving a fractional bundle $x_i$ of cost $c_i(x_i)\le 1$.

\subsection{Rounding Fractional Allocations}\label{sec:rounding}

\begin{figure}[htbp]
\definecolor{slotone}{RGB}{217,232,246}
\definecolor{slottwo}{RGB}{250,225,184}
\definecolor{slotthree}{RGB}{217,237,218}
\definecolor{slotfour}{RGB}{233,221,242}
\definecolor{slotfive}{RGB}{247,218,218}
\definecolor{agentone}{RGB}{38,91,145}
\definecolor{agenttwo}{RGB}{168,101,25}

\begin{subfigure}[t]{.65\textwidth}
\centering
\vspace{0pt}
\begin{tikzpicture}[x=1cm,y=1cm,font=\small,line width=.55pt]
 \path[use as bounding box] (-1.0,1.2) rectangle (8.6,7.8);
 \begin{scope}[xscale=.75]
  \node[anchor=east,text=agentone] at (-.2,7.075) {$x_1$};
  \foreach \l/\r/\col/\e/\f in {
   0/2.4/slotone/1/{3/5},
   2.4/5.2/slottwo/2/{7/10},
   5.2/6.8/slotthree/3/{2/5},
   6.8/10/slotfour/4/{4/5},
   10/11.2/slotfive/5/{3/10}} {
   \filldraw[fill=\col] (\l,6.8) rectangle (\r,7.35);
   \node[anchor=south] at ({(\l+\r)/2},7.42) {$e_{\e}$};
   \node at ({(\l+\r)/2},7.075) {$\f$};
  }

  \node[anchor=east] at (-.2,5.525) {Slots};
  \foreach \l/\r/\col/\e/\f in {
   0/2.4/slotone/1/{3/5},
   2.4/4/slottwo/2/{2/5},
   4/5.2/slottwo/2/{3/10},
   5.2/6.8/slotthree/3/{2/5},
   6.8/8/slotfour/4/{3/10},
   8/10/slotfour/4/{1/2},
   10/11.2/slotfive/5/{3/10}} {
   \filldraw[fill=\col] (\l,5.25) rectangle (\r,5.8);
   \node[anchor=south] at ({(\l+\r)/2},5.87) {$e_{\e}$};
   \node at ({(\l+\r)/2},5.525) {$\f$};
  }
  \foreach \a in {4,8} {
   \draw[dashed,line width=.8pt] (\a,5.18)--(\a,6.8);
  }
  \foreach \a in {2,6,9.6} {
   \draw[-{Stealth[length=1.6mm]}] (\a,6.55)--(\a,6.25);
  }
  \foreach \l/\r/\j in {0/4/1,4/8/2,8/11.2/3} {
   \draw (\l,5.12)--(\l,5.02)--(\r,5.02)--(\r,5.12);
   \node[text=agentone] at ({(\l+\r)/2},4.77) {$S_{1,\j}$};
  }

  \node[anchor=east,text=agenttwo] at (-.2,3.875) {$x_2$};
  \foreach \l/\r/\col/\e/\f in {
   0/1.6/slotone/1/{2/5},
   1.6/2.8/slottwo/2/{3/10},
   2.8/5.2/slotthree/3/{3/5},
   5.2/6/slotfour/4/{1/5},
   6/8.8/slotfive/5/{7/10}} {
   \filldraw[fill=\col] (\l,3.6) rectangle (\r,4.15);
   \node[anchor=south] at ({(\l+\r)/2},4.22) {$e_{\e}$};
   \node at ({(\l+\r)/2},3.875) {$\f$};
  }

  \node[anchor=east] at (-.2,2.325) {Slots};
  \foreach \l/\r/\col/\e/\f in {
   0/1.6/slotone/1/{2/5},
   1.6/2.8/slottwo/2/{3/10},
   2.8/4/slotthree/3/{3/10},
   4/5.2/slotthree/3/{3/10},
   5.2/6/slotfour/4/{1/5},
   6/8/slotfive/5/{1/2},
   8/8.8/slotfive/5/{1/5}} {
   \filldraw[fill=\col] (\l,2.05) rectangle (\r,2.6);
   \node[anchor=south] at ({(\l+\r)/2},2.67) {$e_{\e}$};
   \node at ({(\l+\r)/2},2.325) {$\f$};
  }
  \foreach \a in {4,8} {
   \draw[dashed,line width=.8pt] (\a,1.98)--(\a,3.6);
  }
  \foreach \a in {2,6,8.4} {
   \draw[-{Stealth[length=1.6mm]}] (\a,3.35)--(\a,3.05);
  }
  \foreach \l/\r/\j in {0/4/1,4/8/2,8/8.8/3} {
   \draw (\l,1.92)--(\l,1.82)--(\r,1.82)--(\r,1.92);
   \node[text=agenttwo] at ({(\l+\r)/2},1.57) {$S_{2,\j}$};
  }
 \end{scope}
\end{tikzpicture}
\caption{Unit-slot organization.}
\end{subfigure}%
\hspace{5pt}
\begin{subfigure}[t]{.34\textwidth}
\centering
\vspace{0pt}
\begin{tikzpicture}[x=0.9cm,y=0.9cm,font=\small,
 slotnode/.style={rectangle,rounded corners=1.5pt,draw,line width=.65pt,
   minimum width=.68cm,minimum height=.56cm,inner sep=1.5pt},
 itemnode/.style={circle,draw=black!70,line width=.65pt,
   minimum size=.65cm,inner sep=1pt},
 fraction/.style={inner sep=0pt,font=\footnotesize,anchor=center}]
 \path[use as bounding box] (-2.65,-.1) rectangle (2.65,6.5);
 \node[text=agentone,font=\footnotesize] at (-2.15,6.2) {$x_1$ slots};
 \node[font=\footnotesize] at (0,6.2) {Chores};
 \node[text=agenttwo,font=\footnotesize] at (2.15,6.2) {$x_2$ slots};

 \node[slotnode,draw=agentone,fill=agentone!10] (s11) at (-2.15,5.4) {$S_{1,1}$};
 \node[slotnode,draw=agentone,fill=agentone!10] (s12) at (-2.15,3.0) {$S_{1,2}$};
 \node[slotnode,draw=agentone,fill=agentone!10] (s13) at (-2.15,.6) {$S_{1,3}$};
 \node[slotnode,draw=agenttwo,fill=agenttwo!10] (s21) at (2.15,5.4) {$S_{2,1}$};
 \node[slotnode,draw=agenttwo,fill=agenttwo!10] (s22) at (2.15,3.0) {$S_{2,2}$};
 \node[slotnode,draw=agenttwo,fill=agenttwo!10] (s23) at (2.15,.6) {$S_{2,3}$};
 \node[itemnode,fill=slotone]   (e1) at (0,5.5) {$e_1$};
 \node[itemnode,fill=slottwo]   (e2) at (0,4.25) {$e_2$};
 \node[itemnode,fill=slotthree] (e3) at (0,3.0) {$e_3$};
 \node[itemnode,fill=slotfour]  (e4) at (0,1.75) {$e_4$};
 \node[itemnode,fill=slotfive]  (e5) at (0,.5) {$e_5$};

 \draw[draw=agentone!85,line width=.65pt] (s11)--(e1);
 \draw[draw=agentone!85,line width=.65pt] (s11)--(e2);
 \draw[draw=agentone!85,line width=.65pt] (s12)--(e2);
 \draw[draw=agentone!85,line width=.65pt] (s12)--(e3);
 \draw[draw=agentone!85,line width=.65pt] (s12)--(e4);
 \draw[draw=agentone!85,line width=.65pt] (s13)--(e4);
 \draw[draw=agentone!85,line width=.65pt] (s13)--(e5);
 \draw[draw=agenttwo!85,line width=.65pt] (s21)--(e1);
 \draw[draw=agenttwo!85,line width=.65pt] (s21)--(e2);
 \draw[draw=agenttwo!85,line width=.65pt] (s21)--(e3);
 \draw[draw=agenttwo!85,line width=.65pt] (s22)--(e3);
 \draw[draw=agenttwo!85,line width=.65pt] (s22)--(e4);
 \draw[draw=agenttwo!85,line width=.65pt] (s22)--(e5);
 \draw[draw=agenttwo!85,line width=.65pt] (s23)--(e5);

 \node[fraction] (ws111) at (-1.0673,5.7004) {$3/5$};  
 \node[fraction] (ws112) at (-0.8876,4.9604) {$2/5$};  
 \node[fraction] (ws122) at (-1.0447,4.0425) {$3/10$}; 
 \node[fraction] (ws123) at (-1.0675,3.2300) {$2/5$};  
 \node[fraction] (ws124) at (-0.6447,2.4275) {$3/10$}; 
 \node[fraction] (ws134) at (-1.0476,1.5396) {$1/2$};  
 \node[fraction] (ws135) at (-.9173,.2996) {$3/10$};  
 \node[fraction] (ws211) at (1.0673,5.7004) {$2/5$};   
 \node[fraction] (ws212) at (0.70,4.89) {$3/10$};       
 \node[fraction] (ws213) at (1.43,3.95) {$3/10$};      
 \node[fraction] (ws223) at (0.93675,3.2300) {$3/10$};  
 \node[fraction] (ws224) at (.85,2.49) {$1/5$};        
 \node[fraction] (ws225) at (1.18,1.30) {$1/2$};       
 \node[fraction] (ws235) at (1.0673,.2996) {$1/5$};    
\end{tikzpicture}
\caption{Bipartite graph.}
\end{subfigure}

\caption{Unit-slot organization for two complementary fractional bundles $x_1 = (\frac{3}{5},\frac{7}{10},\frac{2}{5},\frac{4}{5},\frac{3}{10})$ and $x_2=1-x_1$, and the corresponding bipartite graph with edge fractions. 
The decomposition produces four matchings $M_1=\left(1,2,4,3,5,0\right)$, $M_2=\left(1,4,5,2,3,0\right)$, $M_3=\left(2,3,4,1,5,0\right)$ and $M_4=\left(2,3,0,1,4,5\right)$, with coefficients $3/10$, $3/10$, $1/5$ and $1/5$, respectively. The entries follow the slot order $\left(S_{1,1},S_{1,2},S_{1,3},S_{2,1},S_{2,2},S_{2,3}\right)$, and $0$ means the corresponding slot is unmatched.}
\label{fig:general-unit-slots}
\end{figure}
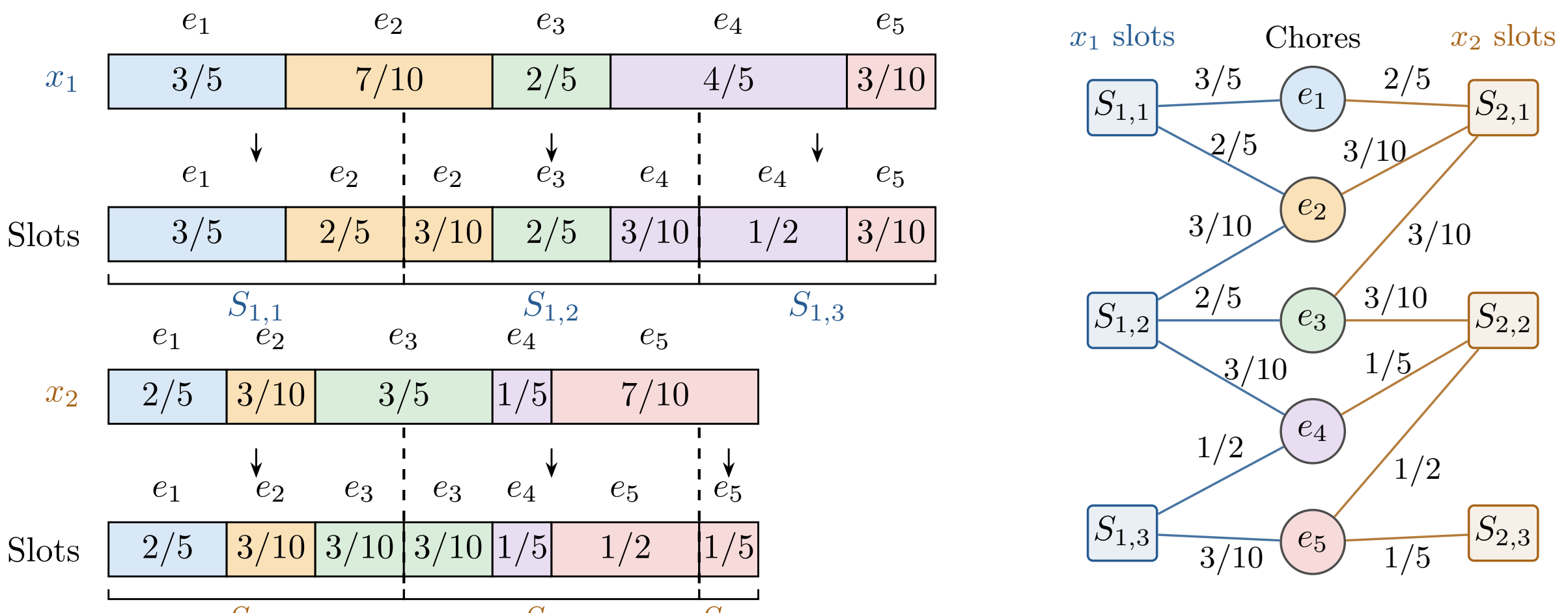

Given a realized vector $\widehat{\eta}$ from Section~\ref{sec:selection}, we now round the resulting complete fractional allocation $x=(x_1,\ldots,x_n)$ into a finite lottery over complete integral allocations.
Recall that for each pure ordinary agent $i$, we have $x_i=x_{i,t}$ for her unique selected bundle index $t$.
Our approach constructs private slots for each agent's fractional bundle $x_i$ and then rounds all agents' slots jointly.
In particular, the slot construction for pure ordinary agents guarantees that they receive an MMS bundle whenever they are assigned no chore from $H$.
To describe the slot construction, we fix an arbitrary agent $i\in N$ and her fractional bundle $x_i$.
\paragraph{General Unit-slot Rounding.}
For a general fractional bundle $x_i$, we list its chore fractions $(x_i(e_1),\ldots,x_i(e_m))$ in nonincreasing order of cost $c_i$ and partition them into consecutive unit slots of total fraction $1$ (except possibly the last), splitting fractions between adjacent slots when necessary.
Let $S_j$ denote the $j$-th slot, let $c_i(S_j)$ be its fractional cost, and define $u_j=\max_{e\in\operatorname{supp}(S_j)} \{c_i(e)\}$.
Since every slot preceding another has total fraction $1$, we have
\begin{equation}\label{eq:unit-slot-estimate}
 u_{j+1}\le\min_{e\in\operatorname{supp}(S_j)} \{c_i(e)\} \le c_i(S_j)
 \quad\text{and hence}\quad
 \sum_{j\ge 2}u_j\le\sum_j c_i(S_j)=c_i(x_i).
\end{equation}
Since $u_1\le\kappa_i$, any integral bundle formed by selecting at most one chore from the support of each slot has cost at most $c_i(x_i)+\kappa_i$ (and at most $c_i(x_i)$ if the chore from $S_1$ is removed), recovering the classic bound of Shmoys and Tardos~\cite{journals/mp/ShmoysT93}.
See Figure~\ref{fig:general-unit-slots}(a) for an illustration.

Let $X_i$ denote the resulting integral bundle.
Since removing the chore from the first slot, if any, reduces the cost to at most $c_i(x_i)\le 1$, the bundle $X_i$ always satisfies MMS1.
To achieve the stronger guarantee of MMS for almost all agents, we refine the slot construction for pure ordinary agents as described next, which ensures that they receive an MMS bundle whenever no chore from $H$ is assigned to them after rounding.

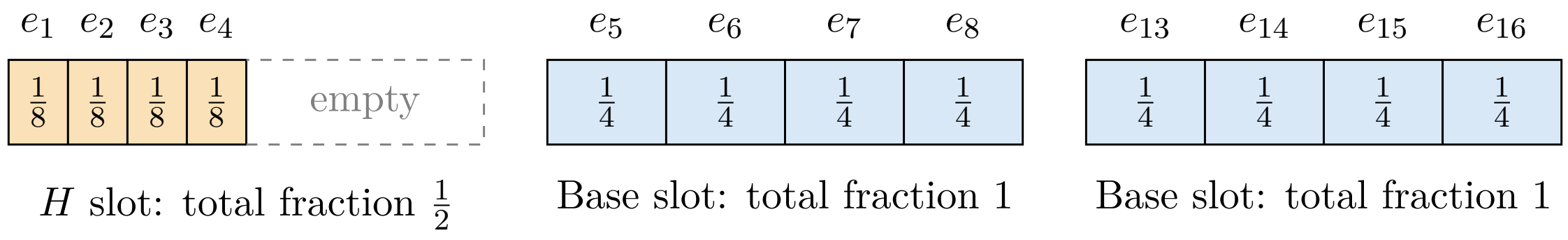
\begin{figure}[htbp]
\centering
\begin{tikzpicture}[x=1cm,y=1cm,font=\normalsize,line width=.55pt]
 \definecolor{sloth}{RGB}{250,225,184}
 \definecolor{slotbase}{RGB}{217,232,246}

 \begin{scope}
  \foreach \j/\e in {0/1,1/2,2/3,3/4} {
   \filldraw[fill=sloth] ({.5625*\j},0) rectangle ({.5625*(\j+1)},.8);
   \node[anchor=south,inner sep=0pt] at ({.5625*(\j+.5)},1) {$e_{\e}$};
   \node[inner sep=0pt] at ({.5625*(\j+.5)},.4) {$\frac18$};
  }
  \draw[dashed,gray] (2.25,0)--(4.5,0)--(4.5,.8)--(2.25,.8);
  \node[text=gray] at (3.375,.4) {empty};
  \node[anchor=north,align=center] at (2.25,-.2) {$H$ slot: total fraction $\frac12$};
 \end{scope}

 \begin{scope}[xshift=5.1cm]
  \foreach \j/\e in {0/5,1/6,2/7,3/8} {
   \filldraw[fill=slotbase] ({1.125*\j},0) rectangle ({1.125*(\j+1)},.8);
   \node[anchor=south,inner sep=0pt] at ({1.125*(\j+.5)},1) {$e_{\e}$};
   \node at ({1.125*(\j+.5)},.4) {$\frac14$};
  }
  \node[anchor=north,align=center] at (2.25,-.2) {Base slot: total fraction $1$};
 \end{scope}

 \begin{scope}[xshift=10.2cm]
  \foreach \j/\e in {0/13,1/14,2/15,3/16} {
   \filldraw[fill=slotbase] ({1.125*\j},0) rectangle ({1.125*(\j+1)},.8);
   \node[anchor=south,inner sep=0pt] at ({1.125*(\j+.5)},1) {$e_{\e}$};
   \node at ({1.125*(\j+.5)},.4) {$\frac14$};
  }
  \node[anchor=north,align=center] at (2.25,-.2) {Base slot: total fraction $1$};
 \end{scope}
\end{tikzpicture}
\caption{Slots for a pure ordinary agent with $m=16$, $H=\left\{e_1,e_2,e_3,e_4\right\}$, $q=4$ and $\tau_{i,t}=1/8$. }
\label{fig:pure-unit-slots}
\end{figure}

\paragraph{Slots for Pure Ordinary Agents.}
Fix a pure ordinary agent $i \in N_o$ and her selected bundle index $t$, so that $x_i = x_{i,t}$ by definition.
We first construct slots for her base bundle $y_{i,t}$ using the original sets $G_{i,t}$ from Section~\ref{sec:fractional-allocations}, distinguishing between two cases based on the index $t$:
\begin{itemize}
    \item If $t < L$, for each block $B_j \subseteq G_{i,t}$, we place the $q$ fractions of size $1/q$ into a single slot.
    Note that each such slot has a total fraction of one, and every chore within it has cost at most $c_i(\hat{e}_j)$, where $\hat{e}_j$ is the representative chore in $B_j$.
    Since the representatives of all blocks in bin $t$ have a total cost of at most one, selecting at most one chore from each slot guarantees that the rounded integral bundle $X_i$ satisfies $c_i(X_i) \leq 1$.
    \item Otherwise (i.e., when $t = L$), we assign a fraction of $1/(q+h)$ to each chore in the original set $G_{i,L}$ and group these fractions into unit slots.
    Their total fractional cost satisfies $c_i(G_{i,L})/(q+h) \le 1$, with each full slot containing $q+h$ chores.
    Recall that in Step~1, we moved the first $\min\{q+h, |G_{i,L}|\}$ chores of $G_{i,L}$ into the excluded set $Z_i$, which correspond precisely to the first slot.
    Hence, the first slot is removed in this case and the sum of the maximum chore costs across the remaining slots is at most one by \eqref{eq:unit-slot-estimate}.
\end{itemize}
We now apply the shift from Section~\ref{sec:fractional-allocations} to these slots: replace each surviving chore $e_{r_j}$ by $e_{K+j}$, deleting its fraction when $K+j>m$.
The resulting slots partition the shifted base bundle $y_{i,t}$.
Since $c_i(e_{K+j})\le c_i(e_{r_j})$, this operation does not increase the maximum chore cost in any slot.
Next, we handle the fractions on the excluded set $H$, whose total fraction in $x_i$ is $K\cdot\tau_{i,t}$.
To incorporate the fractions on $H$, we proceed as follows depending on the value of $K\cdot\tau_{i,t}$:
\begin{itemize}
\item If $K\cdot\tau_{i,t} \le 1$, we place all fractions on $H$ into a single additional slot (see Figure~\ref{fig:pure-unit-slots}).
\item Otherwise (i.e., when $K\cdot\tau_{i,t} > 1$), a single slot cannot place all fractions on $H$. 
We therefore adopt the general unit-slot rounding introduced before. 
Since $H$ consists of the highest-cost chores, the first unit slot is completely filled with fractions of chores from $H$. 
\end{itemize}
Therefore, in both cases, the resulting bundle $X_i$ has cost at most $1+\kappa_i$ and satisfies both MMS1 and $c_i(X_i \setminus H) \le 1$.
In particular, any pure ordinary agent who receives no chore from $H$ is guaranteed to satisfy MMS.

\medskip

\begin{lemma}\label{lem:slots}
There exists a finite lottery over integral allocations $X=(X_1,\ldots,X_n)$ such that:
\begin{itemize}
    \item Every chore is assigned to a single slot, each slot receives at most one chore, full slots always receive a chore, and all slot marginals are preserved.
    \item For every agent $i$ using general unit-slots, the resulting bundle $X_i$ satisfies $c_i(X_i)\le c_i(x_i)+\kappa_i$, and removing the chore in her first slot (if any) leaves cost at most $c_i(x_i)$.
    \item Every pure ordinary agent $i$ receives an MMS1 bundle satisfying $c_i(X_i)\le1+\kappa_i$ and $c_i\!\left(X_i\setminus H\right)\le1$.
\end{itemize}
\end{lemma}
\begin{proof}
Let $x_{i}(e,j)$ denote the fraction of chore $e$ placed in the $j$-th slot $S_{i,j}$ of agent $i$.
Since the slots partition the complete fractional allocation $x$, we have $x_{i}(e,j) \ge 0$ along with
\begin{equation*}
 \sum_{i,j} x_{i}(e,j) = 1 \quad \forall e\in M
 \qquad \text{and} \qquad
 \sum_{e\in M} x_{i}(e,j)\le 1 \quad \forall i, j.
\end{equation*}
Let $\mathcal{S}$ be the collection of all slots across all agents (note that $\vert{}\mathcal{S}\vert{}\ge m$).
To make every slot full, we introduce $\vert{}\mathcal{S}\vert{}-m$ dummy chores, each of total fraction $1$, and greedily fill the slot deficits (splitting dummy chores across slots as needed).
This yields an $\vert{}\mathcal{S}\vert{}\times \vert{}\mathcal{S}\vert{}$ doubly stochastic matrix indexed by all chores (real and dummy) and all slots in $\mathcal{S}$.
By the Birkhoff--von Neumann theorem, this matrix can be written as a finite convex combination of permutation matrices, which equivalently decomposes the bipartite graph between slots and chores into full matchings (see Figure~\ref{fig:general-unit-slots}(b)).
Sampling a permutation matrix according to the convex coefficients and discarding all dummy chores yields the desired lottery: every real chore is assigned to a single slot, each slot receives at most one real chore, and $\Pr[e\text{ is assigned to } S_{i,j}]=x_{i}(e,j)$.
Moreover, any full slot receives no dummy fraction and hence always receives a real chore.
For agents using general unit-slots, the cost bounds follow from~\eqref{eq:unit-slot-estimate}.
For pure ordinary agents, the additional guarantees follow from the slot construction above, since every outcome assigns at most one chore to each slot.
\end{proof}

Therefore, we can sample a matching from this lottery and give each agent the chores matched to her slots, returning the resulting complete integral allocation $\bX=\left(X_1,\ldots,X_n\right)$.

\subsection{The Final Analysis}\label{sec:main-proof}

We now verify the parameter requirements and complete the proof of Theorem~\ref{thm:main}.

\begin{proof}[Proof of Theorem~\ref{thm:main}]
For any outcome $\bX=(X_1,\ldots,X_n)$ of the lottery constructed in Section~\ref{sec:sublinear-disposal}, the slot construction in Step~3 ensures that each agent $i\in N$ receives a bundle $X_i$ of cost at most $1+\kappa_i$ that satisfies MMS1.
Moreover, any pure ordinary agent whose bundle does not satisfy MMS must receive at least one chore from $H$, so there are at most $\vert{}H\vert{}=K$ such agents.
Since there are at most $2(k+1)$ mixed ordinary agents (by Lemma~\ref{lem:reserve-coordination}) and $k$ reserved agents, the total number of agents without an MMS bundle is at most $K+3k+2$.
By our choice of parameters $k$ and $q$, we have $\frac{1}{2}\sqrt{n\log n}<q\le\sqrt{n\log n}$.
For $C_1\ge\sqrt{2}\,C_2$, this yields
\begin{equation*}
 k\ge C_2\cdot n\sqrt{\frac{\log n}{q}}
 \qquad\text{and}\qquad
 K=O(\sqrt{n\cdot q})=O(k).
\end{equation*}
Thus, every supported outcome leaves at most $\tilde{O}(n^{3/4})$ agents without an MMS bundle.

We next establish ex-ante proportionality by showing that the final lottery is equal-marginal on the IDO instance.
For every ordinary agent $i\in N_o$ and chore $e\in M$, Step~2 ensures that $\mathbb{E}[\widehat{\eta}_{i,t}]=\eta_{i,t}$.
Together with Step~1, this yields
\begin{equation*}
 \mathbb{E}[x_i(e)]=\sum_{t=1}^L\eta_{i,t}\cdot x_{i,t}(e)=\frac{1}{n}.
\end{equation*}
Since each reserved agent receives a $1/k$ share of the remaining fraction of every chore before rounding, for all $j\in N_r$ and $e\in M$ we have
\begin{equation*}
 \mathbb{E}[x_j(e)]
 =\frac{1-\sum_{i\in N_o}\mathbb{E}[x_i(e)]}{k}
 =\frac{1-(n-k)/n}{k}=\frac{1}{n}.
\end{equation*}
By Lemma~\ref{lem:slots}, the rounding in Step~3 preserves all marginals, so $\Pr[e\in X_i]=\mathbb{E}[x_i(e)]=1/n$ for all $i\in N$ and $e\in M$.
Thus, all claimed guarantees hold on the IDO instance, and by Lemma~\ref{lem:ido-reduction}, they carry over to the original instance.
\end{proof}

\section{\texorpdfstring{Ex-Ante PROP and Ex-Post $4/3$-MMS}{Ex-Ante PROP and Ex-Post 4/3-MMS}}\label{sec:further-guarantees}\label{sec:four-thirds}

In the preceding sections, guaranteeing exact MMS for almost all agents required partitioning the chores into blocks of size $q$ and coordinating lotteries over $L$ fractional bundles per agent.
When we instead aim for a worst-case $4/3$-MMS guarantee for all agents alongside ex-ante proportionality and ex-post MMS1, a much simpler two-part decomposition suffices.

By adding zero-cost dummy chores if necessary, we may assume that $m\ge 2n$. 
We partition the first $2n$ chores into $n$ \emph{opposite pairs} $Q_1,\ldots,Q_n$ and collect the remaining chores in $V$:
\begin{equation*}
 Q_j=\{e_j,e_{2n+1-j}\} \quad \text{for } j\in[n],
 \qquad\text{and}\qquad
 V=\{e_{2n+1},\ldots,e_m\}.
\end{equation*}
Note that $|Q_j|=2$ for all $j\in[n]$, and the sets $Q_1,\ldots,Q_n,V$ form a partition of $M$.
By the pigeonhole principle and standard MMS reduction properties, we obtain the following.

\begin{lemma}
\label{lem:opposite-pairs}
For every agent $i$ and every $j\in[n]$, we have
\begin{equation*}
 \max_{g\in V}\left\{c_i(g)\right\}\le\frac13
 \quad \text{ and } \quad
 c_i\!\left(Q_j\right)\le\frac43.
\end{equation*}
\end{lemma}
\begin{proof}
We fix an agent $i$.
In any MMS partition for agent $i$, each bundle has cost at most $1$ and therefore contains at most two chores with cost strictly greater than $1/3$.
Thus, at most $2 n$ chores in $M$ have cost greater than $1/3$, which implies $\max_{g\in V}\left\{c_i(g)\right\}\le 1/3$.

We now fix $j\in[n]$.
If $c_i\!\left(e_{2 n+1-j}\right)\le1/3$, then we immediately have $c_i\!\left(Q_j\right)\le1+ 1/3=4/3$.
Otherwise, $c_i\!\left(e_{2 n+1-j}\right)>1/3$, so every chore in the prefix $\left\{e_1,\ldots,e_{2 n+1-j}\right\}$ costs strictly more than $1/3$.
Since $M$ admits a partition into $n$ bundles of cost at most $1$, this prefix can also be partitioned into $n$ bundles of cost at most $1$ (with at most two chores per bundle).
By~\cite[Lemma~6]{journals/teco/BarmanK20}, the minimax-optimal partition of $\left\{e_1,\ldots,e_{2 n+1-j}\right\}$ into $n$ bundles is given by the $j-1$ singletons and $n-j+1$ opposite pairs:
\begin{equation*}
    \left\{e_1\right\},\ldots,\left\{e_{j-1}\right\},
 \left\{e_{j},e_{2 n+1-j}\right\},\ldots,\left\{e_n,e_{n+1}\right\}.
\end{equation*}
Because $Q_j=\left\{e_j,e_{2 n+1-j}\right\}$ is one of the bundles in this partition, we have $c_i\!\left(Q_j\right)\le 1$.
\end{proof}

We next present a randomized completion procedure (Algorithm~\ref{alg:bundle-completion}) for the constructed partition $Q_1,\ldots,Q_n,V$ of $M$, where the chores in $V$
are treated as a single divisible chore during the fractional
completion.
The procedure outputs a fractional allocation in which each agent $i\in N$ receives one integral bundle from $\mathcal{Q} = \{Q_1,\ldots,Q_n\}$ together with a fraction $t_i\in [0,1]$ of every chore in $V$, satisfying $\sum_{i\in N} t_i = 1$.
We write $Q^i\in\mathcal Q$ for the fixed bundle assigned to agent $i$ by the procedure.
The resulting fractional allocation satisfies the following properties:
\begin{itemize}
    \item Each agent $i$'s expected total cost is at most $\PROP_i$, which gives ex-ante PROP.
    \item Any agent $i$ with $t_i > 0$ incurs a total fractional cost of at most $\PROP_i$, which yields an ex-post cost of at most $\PROP_i + 1/3 \le 4/3$ after unit-slot rounding.
    \item Any agent $i$ with $c_i(Q^i) > \PROP_i$ receives $t_i = 0$. Thus, she receives only $Q^i$ after rounding, which costs at most $4/3$ by Lemma~\ref{lem:opposite-pairs}.
\end{itemize}

\paragraph{Algorithm Overview.} 
Our algorithm runs in at most $n$ rounds.
We use $\mathcal{Q}'\subseteq \mathcal{Q}$ and $N'\subseteq N$ to denote the remaining bundles and active agents, respectively.
In each round, one active agent $i\in N'$ is assigned a bundle from $\mathcal{Q}'$ and a fraction $t_i\in [0,1]$ of every chore in $V$.
At a stage with active agents $N'$, let $b$ denote the remaining unassigned fraction of every chore in $V$.
For each active agent $i\in N'$, let $a_i$ be her minimum cost among the remaining bundles in $\mathcal{Q}'$, and let $t_i\in [0,b]$ be the largest fraction of $V$ she can take together with that bundle while keeping her total fractional cost at most $\PROP_i$.
If $\max_{i\in N'}\{t_i\} < b$, we select an agent $i$ maximizing $t_i$ and assign her a cheapest remaining bundle in $\mathcal{Q}'$ (of cost $a_i$) together with fraction $t_i$ of every chore in $V$.
Intuitively, the maximality of $t_i$ ensures that every other remaining agent $j$ evaluates the removed package at least at $\PROP_j$.
Once the set $E=\{i\in N' : t_i=b\}$ is nonempty, we choose an agent $i\in E$ uniformly at random to receive her cheapest remaining bundle in $\mathcal{Q}'$ and the entire remaining fraction $b$ of $V$, and then assign the remaining bundles in $\mathcal{Q}'$ to the remaining agents in $N'$ via a uniformly random bijection (with zero fraction of $V$).

\begin{algorithm}[htbp]
\caption{Completing fixed bundles}\label{alg:bundle-completion}
\begin{algorithmic}[1]
\Require The constructed partition $Q_1,\ldots,Q_n,V$ of $M$ and costs $c_i$ for $i\in N$.
\Ensure A complete fractional allocation with one fixed bundle per agent.
\State Initialize $N' \gets N$, $\mathcal{Q}' \gets \mathcal{Q}$, and $b\gets1$.
\Comment{$b$ is the remaining unassigned fraction}
\While{$N'\neq \emptyset$}
    \State For every active agent $i$, set $a_i\gets\min_{Q\in \mathcal{Q}'} \left\{ c_i(Q) \right\}$ and
    \begin{equation*}
        t_i\gets\max\!\left\{t\in[0,b]:a_i+t\cdot c_i(V)\le\PROP_i\right\}
    \end{equation*}
    \If{$\max_{i\in N'}\{t_i\} < b$}
        \State Choose an arbitrary agent $i\in N'$ maximizing $t_i$.
        \State Give $i$ a bundle $Q^i\in \mathcal{Q}'$ of cost $a_i$ and fraction $t_i$ of every chore in $V$.
        \State Update $N'\gets N'\setminus \{i\}$, $\mathcal{Q}' \gets \mathcal{Q}'\setminus \{Q^i\}$, and $b\gets b-t_i$.
    \Else
        \State Set $E\gets\{i\in N' : t_i=b\}$ and choose $i$ uniformly from $E$.
        \State Give $i$ a bundle $Q^i\in \mathcal{Q}'$ of cost $a_i$ and fraction $t_i=b$ of every chore in $V$.
        \State Remove $i$ from $N'$ and $Q^i$ from $\mathcal{Q}'$.
        \State Give each remaining agent $j\in N'$ a bundle $Q^j\in\mathcal{Q}'$ via a uniformly random bijection.
        \State Set $t_j\gets0$ for every $j\in N'$, and update $N'\gets\varnothing$ and $\mathcal Q'\gets\varnothing$.
    \EndIf
\EndWhile
\State \Return the fractional allocation $x=\left(x_1,\ldots,x_n\right)$.
\end{algorithmic}
\end{algorithm}

Let $x_i$ denote agent $i$'s fractional bundle before rounding: it consists of her assigned integral bundle $Q^i\in \mathcal{Q}$ and fraction $t_i$ of every chore in $V$.

\begin{lemma}\label{lem:fractional-completion}
Algorithm~\ref{alg:bundle-completion} produces a lottery over fractional allocations $x=\left(x_1,\ldots,x_n\right)$ such that $\mathbb E\!\left[c_i(x_i)\right]\le {c_i(M)}/n$.
Moreover, every $x$ in the support has $c_i(x_i)\le c_i(M)/n$ if $t_i > 0$.
\end{lemma}
\begin{proof}
At a stage with $k=|N'|$ active agents, we maintain the invariant that
\begin{equation*}
 \sum_{Q\in\mathcal Q'}c_i(Q)+b\cdot c_i(V)
 \le k\cdot\PROP_i\qquad\forall i\in N'.
\end{equation*}
The invariant holds initially with equality when $N' = N$ and $b=1$.
Furthermore, the invariant implies that $a_i = \min_{Q\in\mathcal{Q}'} \{c_i(Q)\} \le \PROP_i$ (and thus $t_i$ is well-defined) for all $i\in N'$.
Furthermore, when $k=1$, the invariant ensures $a_i+b\cdot c_i(V)\le\PROP_i$, so $t_i=b$ is feasible and the algorithm terminates within $n$ rounds with a complete fractional allocation.
Suppose the invariant holds at the beginning of a round, and agent $i$ is selected in this round.
\begin{itemize}
    \item If $\max_{\ell\in N'}\{t_\ell\}<b$, then $t_i<b$ and agent $i$ receives cost $a_i+t_i\cdot c_i(V)=\PROP_i$.
    By the maximality of $t_i$, every $j\in N'$ satisfies
    $$
    c_j(Q^i) + t_i \cdot c_j(V) \geq a_j+t_i\cdot c_j(V)\ge a_j+t_j\cdot c_j(V)=\PROP_j,$$ which preserves the invariant for the $k-1$ remaining agents.
    \item If $\max_{\ell\in N'}\{t_\ell\} = b$, then agent $i$ is chosen uniformly at random from $E=\{j\in N':t_j=b\}$. Note that every agent $j\notin E$ has $t_j<b$, and hence $a_j+b\cdot c_j(V)>\PROP_j$.
    Since $j\notin E$ receives a uniformly random bundle from the remaining $k-1$ bundles after $i$ removes a bundle of cost at least $a_j$, her expected cost is at most
    \begin{equation*}
        \mathbb E\!\left[c_j(x_j)\right] 
        \leq 
        \frac{k\cdot\PROP_j-(c_j(Q^i)+b\cdot c_j(V))}{k-1}
        \leq 
        \frac{k\cdot\PROP_j-(a_j+b\cdot c_j(V))}{k-1}<\PROP_j.
    \end{equation*}
    Meanwhile, each agent $j\in E$ is chosen as $i$ with probability $1/|E|$ (receiving cost $a_j+b\cdot c_j(V)\leq \PROP_j$) and otherwise receives one of the remaining $k-1$ bundles uniformly at random (whose total cost is at most $k\cdot\PROP_j-(a_j+b\cdot c_j(V))$).
    For $j\in E$, we have
    \begin{align*}
     \mathbb E\!\left[c_j(x_j)\right]
     &\le\frac{a_j+b\cdot c_j(V)}{|E|}+\frac{|E|-1}{|E|}\cdot\frac{k\cdot\PROP_j-(a_j+b\cdot c_j(V))}{k-1}\\
     &=\PROP_j+\frac{k-|E|}{|E|\cdot(k-1)}\cdot\left(a_j+b\cdot c_j(V)-\PROP_j\right)
     \le\PROP_j.
    \end{align*}
    Note that this bound holds even when $|E|=1$.
\end{itemize}
Therefore, we have $\mathbb E\!\left[c_i(x_i)\right]\le {c_i(M)}/n$ for all $i\in N$, and $c_i(x_i)\le c_i(M)/n$ whenever $t_i > 0$.
\end{proof}

We now keep each agent $i$'s assigned bundle $Q^i\in\mathcal{Q}$ fixed and round only the fractional shares of $V$ using the unit-slots of Lemma~\ref{lem:slots}.
Since unit-slot rounding preserves all marginal probabilities, we obtain the following guarantee.

\begin{lemma}\label{lem:bundle-completion}
For the constructed opposite pairs $Q_1,\ldots,Q_n$ and remaining chores $V$, rounding the fractional lottery from Algorithm~\ref{alg:bundle-completion} via Lemma~\ref{lem:slots} produces a finite lottery over MMS1 allocations $\bX=\left(X_1,\ldots,X_n\right)$ that is ex-ante PROP.
In every supported outcome, each agent $i$ either receives only her fixed bundle $Q^i\in\mathcal Q$, or there exists a chore $e\in X_i\cap V$ such that $c_i(X_i\setminus\{e\})\leq c_i(M)/n$.
\end{lemma}

Combining Lemma~\ref{lem:opposite-pairs} and Lemma~\ref{lem:bundle-completion} yields our $4/3$-MMS theorem.

\begin{theorem}\label{thm:four-thirds}
Every instance admits a finite lottery over MMS1 allocations $\bX=\left(X_1,\ldots,X_n\right)$ that is ex-ante PROP and ex-post $4/3$-MMS.
\end{theorem}
\begin{proof}
For IDO instances, combining Lemmas~\ref{lem:opposite-pairs} and~\ref{lem:bundle-completion} with the fact that $\PROP_i\le 1$ directly yields a finite lottery over MMS1 allocations that is both ex-ante proportional and ex-post $4/3$-MMS.
Removing any zero-cost dummy chores and applying Lemma~\ref{lem:ido-reduction} transfers the lottery to the original non-IDO instance while preserving all ex-ante and ex-post guarantees.
\end{proof}

\section*{Declaration for the Use of AI}
We used OpenAI's GPT-6 Astra and GPT-6.1 Sol to assist with mathematical analysis, language editing, notation consistency checks, and LaTeX formatting. 
We began this project in early 2026 with the goal of beating $n/4$ for the subsidy required to achieve MMS for the allocation of chores, and we obtained preliminary results.
GPT-6 Astra helped strengthen the result to sublinear and generalize it to support sublinear disposal and ex-ante PROP.
Through intensive discussion and idea exchange with GPT, we substantially improved the results and simplified the proofs.
All mathematical statements, proofs, calculations, and citations were reviewed and verified by the authors, who take full responsibility for the content of the paper.

\clearpage
\bibliographystyle{alpha}
\bibliography{MMS}
\clearpage
\appendix

\section{Missing Proofs from Section 3}\label{app:deferred-proofs}\label{app:section-three}

In this appendix, we prove the two main technical lemmas from Section~3: the representative-chore packing lemma (Lemma~\ref{lem:core-reduction}) used in Section~\ref{sec:fractional-allocations}, and the joint bundle-selection lemma (Lemma~\ref{lem:reserve-coordination}) used in Section~\ref{sec:selection}.

\subsection{Proof of Lemma~\ref{lem:core-reduction}}\label{app:few-excluded-items}\label{app:core-reduction}

Fix an ordinary agent $i\in N_o$.
As in Section~3, we assume that $n$ is sufficiently large, which ensures that $q$, $L$, and all bin counts throughout the proof are also suitably large.
Our goal is to pack the representatives $\hat e_2,\ldots,\hat e_{m/q}$ into $L$ unit-capacity bins and, when $h>0$, one additional bin of capacity $h/q$, while excluding at most $O(\sqrt L)$ representatives.
While one can easily guarantee at most $O(L)$ excluded representatives, sharpening this bound to $O(\sqrt{L})$ is the key to achieving our sublinear guarantee.
Moreover, to ensure that the vectors $\left(x_{i,t}(e)\right)_t$ change value at most $O(L^3)$ times along the chore order (which bounds the number of distinct chore-fraction vectors in the union bound of Section~\ref{sec:selection} independently of $m$), the assignment of representatives to bins or to the excluded set must switch at most $O(L^3)$ times along the sequence $\hat e_2,\ldots,\hat e_{m/q}$.

The proof proceeds in two stages.
\begin{itemize}
    \item First, in Appendices~\ref{app:repeated-halving} and~\ref{app:halving-step}, we start from an MMS partition for agent $i$ and apply $\ell=\log_2 q$ rounds of repeated halving to pack $\hat e_2,\ldots,\hat e_{m/q}$ into the target bin capacities while excluding at most $O(\sqrt L)$ representatives.
    \item Second, in Appendix~\ref{app:rearranging}, we rearrange this packing via a minimum-cost linear program and cycle rounding so that the bin and exclusion labels change at most $O(L^3)$ times along the representative order, while excluding at most $O(\sqrt L)$ additional representatives.
\end{itemize}

\subsubsection{Repeated Halving and the Main Invariant}\label{app:repeated-halving}

\paragraph{Initial $q$-Copy Packing.}
Fix an MMS partition of $M$ for agent $i$ into $n$ bundles of cost at most $1$.
For every chore $e$ in this partition, if $e\in B_t$ with $t<m/q$, we replace $e$ with a copy of $\hat e_{t+1}$ (the representative of the next block $B_{t+1}$), whereas if $e\in B_{m/q}$, we discard $e$.
Since $\hat e_{t+1}$ follows every chore of $B_t$ in the IDO ordering, this replacement never increases the cost of any bundle.
Because every block $B_t$ has size $q$, this construction packs $q$ copies of every representative $\hat e_2,\ldots,\hat e_{m/q}$ into $n$ unit-capacity bins.

\paragraph{Marked Copies.}
Recall that $q=2^\ell$ is a power of $2$.
From this initial $q$-copy packing, we perform $\ell=\log_2 q$ halving steps.
Every step retains half the copies of every representative and merges pairs of bins into new bins whose capacity is the average of the two original capacities.
To keep the bin loads within capacity, the procedure may \emph{mark} a small number of copies, where a marked copy contributes $0$ to its bin's load and we refer to the total cost of the unmarked copies in a bin as its \emph{load}.
We write $M_j$ for the number of marked copies after $j$ steps.

\medskip

\begin{invariant}\label{inv:halving}
After $j\in\{0,\ldots,\ell\}$ halving steps, every representative $\hat e_2,\ldots,\hat e_{m/q}$ has exactly $q/2^j$ copies (both marked and unmarked), which are packed into $\lfloor n/2^j\rfloor$ unit-capacity bins and one additional bin of capacity $r_j=n/2^j-\lfloor n/2^j\rfloor$ if $r_j>0$.
Moreover, every bin's load is at most its capacity, and $M_j\le 8 C_3\sqrt{n/2^j}$ for a constant $C_3>0$.
\end{invariant}

For $j=0$ (before any halving steps), the initial $q$-copy packing satisfies Invariant~\ref{inv:halving} with $M_0=0$, $n$ unit-capacity bins, and $r_0=0$.
In Appendix~\ref{app:halving-step}, we prove that every halving step preserves Invariant~\ref{inv:halving} from $j$ to $j+1$.
After all $\ell=\log_2 q$ steps, every representative $\hat e_2,\ldots,\hat e_{m/q}$ has $q/2^\ell=1$ copy, packed into $L=\lfloor n/q\rfloor$ unit-capacity bins and one additional bin of capacity $r_\ell=h/q$ when $h>0$.
The number of marked copies is $M_\ell\le 8 C_3\sqrt{n/q}=O(\sqrt L)$.
We then exclude the $M_\ell$ representatives whose single remaining copy is marked, which leaves only unmarked copies and yields a feasible packing under agent $i$'s original costs.

\subsubsection{A Halving Step}\label{app:halving-step}

In this section, we focus on the transition from step $j$ to step $j+1$.
Suppose the invariant holds at the end of step $j$.
We show that the invariant is maintained at the end of step $j+1$.
By the invariant, there are $\lfloor n/2^j\rfloor$ unit-capacity bins and one additional bin of capacity $r_j=\frac{n}{2^j}-\left\lfloor\frac{n}{2^j}\right\rfloor$, if $r_j > 0$.
Note that each halving step merges the bins in pairs, so their total number must be even.
We add an empty bin of capacity $0$ when necessary, as follows:
\begin{itemize}
\item If $\lfloor n/2^j\rfloor$ is even, we add an empty bin of capacity $0$ and pair it with the bin of capacity $r_j$ when $r_j>0$, and we add no bin when $r_j=0$.
\item If $\lfloor n/2^j\rfloor$ is odd, we select an arbitrary unit-capacity bin.
When $r_j>0$, we pair it with the existing bin of capacity $r_j$.
Otherwise ($r_j=0$), we add an empty bin of capacity $0$ and pair it with the selected unit-capacity bin.
\end{itemize}
Therefore, we assume that there is an even number $b$ of bins with capacities $T_1,\ldots,T_b$.
There are $q/2^j$ copies of each chore, and $M_j\le 8 C_3\cdot\sqrt{n/2^j}$ copies are marked.

\begin{lemma}
\label{lem:single-round-halving}
There exists a selection retaining exactly half the copies of each representative and a pairing of the $b$ bins that keeps the pair involving a fractional-capacity or padding bin together, if present, such that, after marking additional copies, every merged pair $\left(u,v\right)$ has retained load at most $\left(T_u+T_v\right)/2$ and the total number $M_{j+1}$ of marked retained copies satisfies $M_{j+1} \le {M_j}/{2}+C_3\cdot\sqrt b$.
\end{lemma}

Before we prove Lemma~\ref{lem:single-round-halving}, we verify that this step maintains Invariant~\ref{inv:halving}.
First, the procedure retains $q/2^{j+1}$ copies of every representative.
Moreover, the merged unit-capacity pairs form $\lfloor n/2^{j+1}\rfloor$ bins of capacity $1$, and the pair that includes the bin of capacity $r_j$, if present, forms a bin of capacity
\begin{equation*}
 r_{j+1}=
 \begin{cases}
 r_j/2,&\text{if }\lfloor n/2^j\rfloor\text{ is even},\\
 \left(1+r_j\right)/2,&\text{if }\lfloor n/2^j\rfloor\text{ is odd},
 \end{cases}
\end{equation*}
These match the integer and fractional parts of $n/2^{j+1}$, and every merged bin's load is at most its capacity.
Finally, the total number of marked retained copies satisfies
\begin{equation*}
 M_{j+1}\le\frac{M_j}{2}+C_3\cdot \sqrt{\left\lceil n/2^j\right\rceil+1}
 \le\left(4+\sqrt2\right)\cdot C_3\cdot \sqrt{\frac n{2^j}}
 \le 8 C_3\cdot \sqrt{\frac n{2^{j+1}}}.
\end{equation*}
Thus, Invariant~\ref{inv:halving} holds after $j+1$ steps.
It remains to prove Lemma~\ref{lem:single-round-halving}.

\begin{proof}[Proof of Lemma~\ref{lem:single-round-halving}]
We arbitrarily pair the copies of every representative and join every pair by a \emph{chore-edge}.
Our procedure will retain exactly one endpoint of every chore-edge.
Within every bin, we sort the unmarked copies in non-increasing order of cost and place the marked copies at the end with cost $0$.
We then join consecutive copies in disjoint pairs by \emph{local-edges}, and leave the last copy unpaired if the bin contains an odd number of copies.
Note that the two endpoints of a local-edge need not be copies of the same representative.

The multigraph formed by these two matchings decomposes into alternating paths and even cycles, and is therefore bipartite.
Consequently, we can color the copies red and blue so that every chore-edge and every local-edge has one red endpoint and one blue endpoint.
We next label the chore-edges $1,2,\ldots$ by traversing every path from one end and every cycle from an arbitrary starting vertex, where we complete the labels of one component before we move to the next.
We write $t$ for the label of a chore-edge.
By construction, every local-edge connects copies whose chore-edges carry consecutive labels within the same component, except for a single closing local-edge in every cycle.
See Figure~\ref{fig:item-and-local-pairs}(a) for an illustration of the chore-edges, local-edges, and chore-edge labels.

\begin{figure}[ht]
\centering
\begin{subfigure}[t]{0.5\textwidth}
\centering
\begin{tikzpicture}[
    x=1cm, y=1.2cm,
    bin/.style={
        draw=black!65, rounded corners=2pt, line width=0.5pt
    },
    pair edge/.style={draw=black, line width=0.65pt},
    local edge/.style={draw=black, line width=0.65pt},
    copy/.style={
        circle, minimum size=7mm, inner sep=0pt, font=\scriptsize
    },
    red copy/.style={copy, fill=red!28},
    blue copy/.style={copy, fill=blue!25},
    kept/.style={draw=black, line width=1.05pt},
    not kept/.style={draw=black!35, line width=0.35pt},
    pair label/.style={
        fill=white, inner sep=1pt, font=\scriptsize
    }
]

\foreach \x/\k in {0/1,2/2,4/3,6/4} {
    \draw[bin] (\x-0.58,0.45) rectangle (\x+0.58,3.75);
    \node[font=\small,above] at (\x,3.75) {Bin $\k$};

    \draw[local edge] (\x,3.3)
        .. controls (\x-0.60,3.3) and (\x-0.60,2.5) ..
        (\x,2.5);
    \draw[local edge] (\x,1.7)
        .. controls (\x-0.60,1.7) and (\x-0.60,0.9) ..
        (\x,0.9);
}

\draw[pair edge] (0,3.3)
    -- node[above,pair label] {$1$} (2,3.3);
\draw[pair edge] (2,2.5)
    -- node[above,pair label] {$2$} (4,2.5);
\draw[pair edge] (4,3.3)
    -- node[above,pair label] {$3$} (6,3.3);
\draw[pair edge] (6,2.5) -- (6,1.7);
\node[pair label] at (6.32,2.1) {$4$};

\draw[pair edge] (6,0.9)
    -- node[pos=0.5,above,pair label] {$5$} (4,1.7);
\draw[pair edge] (2,1.7)
    -- node[pos=0.5,above,pair label] {$6$} (4,0.9);
\draw[pair edge] (2,0.9)
    -- node[below,pair label] {$7$} (0,0.9);
\draw[pair edge] (0,1.7) -- (0,2.5);
\node[pair label] at (0.32,2.1) {$8$};

\node[red copy,kept]      at (0,3.3) {$\hat e_2$};
\node[blue copy,not kept] at (0,2.5) {$\hat e_3$};
\node[red copy,kept]      at (0,1.7) {$\hat e_3$};
\node[blue copy,not kept] at (0,0.9) {$\hat e_5$};

\node[blue copy,not kept] at (2,3.3) {$\hat e_2$};
\node[red copy,not kept]  at (2,2.5) {$\hat e_4$};
\node[blue copy,not kept] at (2,1.7) {$\hat e_5$};
\node[red copy,kept]      at (2,0.9) {$\hat e_5$};

\node[red copy,not kept] at (4,3.3) {$\hat e_2$};
\node[blue copy,kept]    at (4,2.5) {$\hat e_4$};
\node[blue copy,kept]    at (4,1.7) {$\hat e_4$};
\node[red copy,kept]     at (4,0.9) {$\hat e_5$};

\node[blue copy,kept]     at (6,3.3) {$\hat e_2$};
\node[red copy,not kept]  at (6,2.5) {$\hat e_3$};
\node[blue copy,kept]     at (6,1.7) {$\hat e_3$};
\node[red copy,not kept]  at (6,0.9) {$\hat e_4$};

\end{tikzpicture}
\caption{The edges and the retained copies.}
\end{subfigure}
\hspace{5pt}
\begin{subfigure}[t]{0.4\textwidth}
\centering
\begin{tikzpicture}[x=0.7cm,y=0.15cm]

\draw[->] (0,-5.5) -- (0,18) node[above] {$p(t)$};
\draw[->] (0,0) -- (8.5,0) node[right] {$t$};
\node[font=\scriptsize,below left] at (0,0) {$0$};

\draw[thick,domain=0:8,samples=101]
    plot (\x,{(\x-1.5)*(\x-5.5)});

\foreach \t in {1,...,8} {
    \draw (\t,0.5) -- (\t,-0.5);
    \node[font=\scriptsize,below] at (\t,-0.6) {$\t$};
}

\foreach \t in {1,6,7,8} {
    \fill[red!75!black]
        (\t,{(\t-1.5)*(\t-5.5)}) circle[radius=1.7pt];
}
\foreach \t in {2,3,4,5} {
    \fill[blue!75!black]
        (\t,{(\t-1.5)*(\t-5.5)}) circle[radius=1.7pt];
}

\end{tikzpicture}
\caption{$p(t)=\left(t-3/2\right)\cdot\left(t-11/2\right)$ on $\left[0,8\right]$.}
\end{subfigure}

\caption{
Straight edges are chore-edges, and curved edges are local-edges.
The numbers label the chore-edges in their order along the alternating cycle.
A thick outline indicates a retained copy: red is retained when $p(t)>0$, and blue when $p(t)<0$.
}
\label{fig:item-and-local-pairs}
\end{figure}
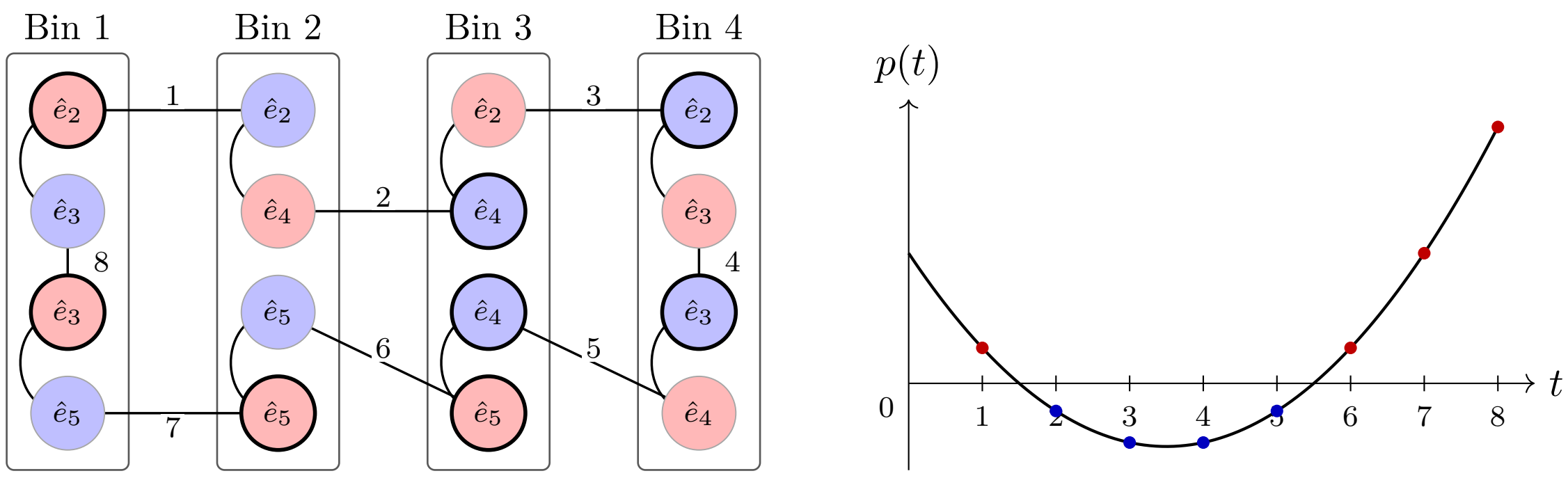

Next, we discuss how to choose one color for each chore-edge (to retain half the copies of each representative) and pair the bins so that a small number of new markings suffice to prove Lemma~\ref{lem:single-round-halving}.
Note that by choosing one color for all chore-edges to retain, pairing unit-capacity bins arbitrarily and marking at most one copy per merged bin, we can halve the copies of each representative while maintaining the capacity requirement.
However, this may require $O(b)$ new markings, which does not meet the $O(\sqrt b)$ requirement.
Therefore, we use the Borsuk--Ulam theorem to choose the colors so that most bins can be paired within their capacity bounds after a few markings.
For the remaining bins, the local-edges allow us to reduce their loads with a few more markings.
Altogether, $O(\sqrt b)$ new markings suffice.

For each copy $s$, we denote its effective cost before the halving step by $w_s$ and the label of its chore-edge by $t_s$.
Here $w_s=0$ if $s$ is already marked.
To guide the selection of colors for each chore-edge, we introduce a variable $\alpha_t\in\left[-1,1\right]$ for each chore-edge $t$, representing that we choose $\left(1+\alpha_t\right)/2$ of the red endpoint and $\left(1-\alpha_t\right)/2$ of the blue endpoint.
The case when $\alpha_t\in\left\{-1,1\right\}$ corresponds to selecting exactly one color for the chore-edge $t$.
Given $\alpha_t\in\left[-1,1\right]$ for the chore-edges $t$, the resulting fractional load of bin $v$ is
\begin{equation*}
 \frac12\cdot\sum_{s\text{ in bin }v}w_s-\frac{D_v(\alpha)}2
 \le\frac{T_v-D_v(\alpha)}2,
 \quad\text{where}\quad
 D_v(\alpha)=
 \sum_{\substack{s\text{ blue}\\\text{in bin }v}}
 w_s\cdot\alpha_{t_s}
 -
 \sum_{\substack{s\text{ red}\\\text{in bin }v}}
 w_s\cdot\alpha_{t_s}.
\end{equation*}

\paragraph{Step 1: Fixing $\alpha$.}
Set $d=\left\lceil\sqrt b\right\rceil$.
Set aside the pair involving a fractional-capacity or padding bin, if present; otherwise, set aside any two unit-capacity bins as a pair.
This leaves $2 b'$ unit-capacity bins, where $b'=b/2-1$.
We claim that there exist a nonzero polynomial $p$ of degree at most $d$ and a vector $\alpha$ with $\alpha_t\in\left[-1,1\right]$ such that $\alpha_t=\operatorname{sgn}\left(p(t)\right)$ when $p(t)\ne0$, and the sorted values among the remaining $2 b'$ unit-capacity bins
$D_{(1)}(\alpha)\ge\cdots\ge D_{\left(2 b'\right)}(\alpha)$ satisfy
\begin{equation}\label{eq:short-halving-symmetry}
 D_{(k_a)}(\alpha)+D_{\left(2 b'+1-k_a\right)}(\alpha)=0,
 \qquad
 k_a=\left\lceil\frac{a\cdot b'}d\right\rceil,
 \qquad a\in[d].
\end{equation}
To prove this, for $\varepsilon>0$ and each coefficient vector $z=\left(z_0,\ldots,z_d\right)$ on the unit sphere
$\mathbb S^d=\left\{z\in\mathbb R^{d+1}:\|z\|_2=1\right\}$, define
\begin{equation*}
 p_z(t)=\sum_{a=0}^d z_a\cdot t^a,
 \qquad
 \alpha_t^\varepsilon(z)=
 \max\left\{-1,\min\left\{1,p_z(t)/\varepsilon\right\}\right\}.
\end{equation*}
For fixed $\varepsilon$, every coordinate of $\alpha^\varepsilon(z)$ is continuous in $z$.
Moreover, replacing $z$ by $-z$ negates the polynomial and swaps the red and blue preferences, so
$\alpha^\varepsilon\left(-z\right)=-\alpha^\varepsilon(z)$.
Consequently, every bin difference is negated as well.
Sorting preserves continuity: when two bins exchange ranks, their differences meet, so the sorted values do not jump.
Negating all differences reverses their sorted order, giving
$D_{(r)}\left(-\alpha\right)=-D_{\left(2 b'+1-r\right)}(\alpha)$.
Thus, although the bins occupying particular ranks may change, replacing $z$ by $-z$ negates each sum in~\eqref{eq:short-halving-symmetry}.
We can now apply the Borsuk--Ulam theorem~\cite{journals/mss/SimmonsS03}.
The form we use says that if $d$ continuous quantities depend on a point of $\mathbb S^d$, and all change sign when that point is replaced by its opposite, then some point makes all $d$ quantities zero simultaneously.
Here the point is the normalized coefficient vector $z$, and the quantities are exactly the $d$ sums in~\eqref{eq:short-halving-symmetry}.
The continuity and sign-reversal properties verified above therefore give a vector $z^\varepsilon\in\mathbb S^d$ satisfying all these equations.
By compactness, as $\varepsilon\to0$, there is a subsequence along which
$\left(z^\varepsilon,\alpha^\varepsilon\left(z^\varepsilon\right)\right)$
converges to a limit $\left(z,\alpha\right)$.
Since $\|z\|_2=1$, the limit polynomial $p=p_z$ is nonzero.
Whenever $p(t)\ne0$, the approximating polynomial values eventually have the same sign and magnitude greater than $\varepsilon$, so
$\alpha_t=\operatorname{sgn}\left(p(t)\right)$.
By continuity,~\eqref{eq:short-halving-symmetry} continues to hold at $\alpha$, establishing the claim.

\paragraph{Step 2: Retaining and Marking Copies.}
For each chore-edge $t$, retain the red copy if $p(t)>0$, the blue copy if $p(t)<0$, and either copy if $p(t)=0$.
Between this selection and its complement (which takes the opposite copy in every chore-edge and corresponds to replacing $\left(p,\alpha\right)$ with $\left(-p,-\alpha\right)$), choose one that retains at most $M_j/2$ already-marked copies.
Because $p$ is a nonzero polynomial of degree at most $d$, it has at most $d$ real roots.
First, mark the retained copy in every chore-edge $t$ with $p(t)=0$; this adds at most $d$ marks.
The retained load of every bin $v$ is at most $\left(T_v-D_v(\alpha)\right)/2$.
Second, whenever a local-edge has two retained copies of positive unmarked cost, mark a copy of maximum cost.
Such a local-edge must join two chore-edges where $p$ takes opposite nonzero signs, so $p$ has a root strictly between their labels.
The label intervals of the non-closing local-edges have disjoint interiors, so at most $d$ of these edges require a mark.
Similarly, a closing local-edge in a cycle requires a mark only if $p$ has a root in the label interval of that cycle.
These cycle intervals are disjoint, so at most $d$ closing local-edges require a mark.
This step therefore adds at most $2 d$ marks and ensures that every local-edge has at most one retained endpoint of positive unmarked cost.

\paragraph{Step 3: Pairing the Bins.}
Let $g=\left\lceil b'/d\right\rceil$.
Among the $2 b'$ unit-capacity bins, set aside the $2\left\lfloor g/2\right\rfloor$ bins with the smallest $D_v(\alpha)$, leaving an even number of bins.
Pair the remaining bins of ranks $r$ and $2 b'-2\cdot\left\lfloor g/2\right\rfloor+1-r$ for $r=1,\ldots,b'-\left\lfloor g/2\right\rfloor$.
The sampled ranks $k_a$ have gaps at most $g$, with $k_1=g$ and $k_d=b'$.
Thus, for each paired rank $r$, we can choose $a\in[d]$ such that
\begin{equation*}
 r\le k_a\le r+g-1
 \le r+2\cdot\left\lfloor g/2\right\rfloor.
\end{equation*}
For the resulting pair of bins $u$ and $v$, the sorted order and~\eqref{eq:short-halving-symmetry} therefore give $D_u(\alpha)+D_v(\alpha)\ge0$.
Hence the merged unit-capacity bin has load at most $\left(T_u+T_v\right)/2=1$.
Pair the $2\cdot\left\lfloor g/2\right\rfloor$ set-aside unit-capacity bins arbitrarily, and include the pair set aside in Step~1; let $U$ be the set of at most $g+2$ bins in these final pairs.
In each bin $v\in U$, mark its first retained copy of positive unmarked cost in the local order, if any; this is a costliest currently unmarked retained copy, and adds at most $\left|U\right|$ marks in total.
Every other retained copy of positive unmarked cost can be matched to a distinct unretained or newly marked copy in an earlier local-edge, whose original effective cost is at least as large.
Thus the load after this last marking is at most half the original load, and hence at most $T_v/2$.
Every merged pair $\left(u,v\right)$ in $U$ therefore has retained load at most $\left(T_u+T_v\right)/2$.
In total, the number of marked retained copies is
\begin{equation*}
 M_{j+1}
 \le\frac{M_j}{2}+d+2 d+\left|U\right|
 \le\frac{M_j}{2}+3 d+\frac bd+4
 \le\frac{M_j}{2}+C_3\cdot\sqrt b,
\end{equation*}
where the last inequality holds for all $C_3>5$ and sufficiently large $b$.
\end{proof}

\subsubsection{Rearranging the Packing into Few Segments}\label{app:rearranging}

After the $\ell$ halving steps, exclude the $M_\ell=O(\sqrt L)$ representatives whose remaining copies are marked.
By Invariant~\ref{inv:halving}, the remaining representatives fit into $L$ unit-capacity bins and, if $h>0$, one fractional-capacity bin of capacity $h/q$.
Write $b\in \{ L, L+1 \}$ for the number of bins, keep their capacities $T_1,\ldots,T_b$, and set $d=\lceil\sqrt b\rceil$.
We want to rearrange the representatives so that nearby chores usually go to the same bin. We first leave some room within each bin for adjustments, then reorganize the packing without increasing its loads. Finally, we fix the errors caused by rounding, using the room we created and excluding a few more representatives.

\paragraph{Step 1: Creating Slack.}
Exclude up to $2 d$ of the costliest remaining representatives.
Let $\kappa$ be the maximum remaining cost, let $T'_v$ be bin $v$'s remaining load, and let $\mathcal V$ contain the bins from which representatives were just excluded.
Then we have $\left|\mathcal V\right|\le 2 d$ and $\sum_{v\in\mathcal V}\left(T_v-T'_v\right)\ge 2 d\cdot\kappa$.
Partition the remaining costs (which lie in $(0,\kappa]$) into the ranges
\begin{equation*}
 I_k=\left(\frac{\left(k-1\right)\cdot\kappa}{d},\frac{k\cdot\kappa}{d}\right],
 \qquad k\in[d].
\end{equation*}
For each $I_k$, keep up to $b(b-1)$ of its cheapest representatives \emph{fixed} in their current bins, taking all of them when fewer are available.
Call the others \emph{movable}, and let $b_{vk}$ be the number of movable representatives of cost $a\in I_k$ assigned to bin $v$.

\paragraph{Step 2: A Structured Fractional Packing.}
Recall that we assumed that the costs are distinct.
Each movable cost $a$ identifies one representative.
Let $z_{v,a}\ge 0$ be its fraction assigned to bin $v$, and choose an optimal basic solution $z^*$ to
\begin{align*}
 \text{minimize}\quad &\sum_{v,a}v\cdot a^2\cdot z_{v,a}\\
 \text{subject to}\quad
 &\sum_{v=1}^b z_{v,a}=1 &&\forall a,\\
 &\sum_{a\in I_k}z_{v,a}=b_{vk} &&\forall v,k,\\
 &\sum_a a\cdot z_{v,a}
   +\sum_{\substack{\text{fixed }\hat e_t\\\text{in bin }v}}c_i\!\left(\hat e_t\right)=T'_v
 &&\forall v.
\end{align*}
The original packing makes these constraints feasible.
All sums over $a$ involve only movable representatives.
Optimality of $z^*$ gives a useful order property: if $a_1<a_2<a_3$ belong to one range and $z^*_{u,a_1},z^*_{v,a_2},z^*_{u,a_3}>0$ for $u\ne v$, then $u<v$.
Indeed, if $u>v$, then moving a small fraction of the middle representative from $v$ to $u$ and moving the same total fraction back using the two endpoint representatives, in proportions $\frac{a_3-a_2}{a_3-a_1}$ and $\frac{a_2-a_1}{a_3-a_1}$, strictly decreases the sum of squares objective, which is a contradiction.
In particular, two bins cannot alternate along four increasing costs in the support of $z^*$.

\paragraph{Step 3: Rounding the Packing.}
For each $I_k$, form a bipartite graph between the bins and movable costs $a\in I_k$, using the edges with $0<z^*_{v,a}<1$.
We use the cycle-rounding mechanism of Ageev and Sviridenko~\cite[Section~2]{journals/jco/AgeevS04}, which preserves the supply and count constraints, keeps integral assignments fixed, and introduces no new assignments outside the support of $z^*$.
We choose the direction of each cycle update as described below to control the changes in bin loads.
There are at most $b-1$ independent cycles across these graphs.
Indeed, shifts around cycles preserve all supplies and counts, while the bin-load equations impose at most $b-1$ additional independent conditions because total cost is conserved.
If more independent cycle directions existed, a nonzero combination would preserve all constraints, contradicting that $z^*$ is basic.
Each cycle update makes at least one edge integral and removes it from the fractional graph, reducing the number of independent cycles.
Thus at most $b-1$ updates suffice to obtain an integral assignment.
To choose the direction of an update, consider its cycle and view each bin as a connection between its two neighboring costs.
By the order property from Step~2, two connections cannot interleave in cost order.
Hence the cycle visits the costs in increasing order, with the last bin connecting the largest cost back to the smallest.
Indeed, any connection other than these would leave some costs strictly between its endpoints and some outside them; the rest of the cycle would have to connect these two groups, creating an interleaving connection.
Choose the rounding direction so that the last bin gains load and every other bin replaces higher costs with lower costs.
The amount shifted is less than one, and all costs lie in a range of width $\kappa/d$.
Consequently, the increase in the last bin and the combined decrease in the other bins are both at most $\kappa/d$.

\paragraph{Step 4: Repairing the Packing.}
For each $I_k$ containing movable representatives, some bin contains at least $b-1$ of its fixed representatives.
Select one such bin $v_k$ for each range and add these bins to $\mathcal V$.
Since there are $d$ ranges, we have $\left|\mathcal V\right|\le 3 d$.
For $v\notin\mathcal V$, let $u_{vk}$ count the cycle updates that increased bin $v$'s load using costs in $I_k$.
Swap its $\min\left\{u_{vk},b_{vk}\right\}$ costliest movable representatives of cost in $I_k$ with that many fixed representatives from $v_k$.
Since each update increases only one bin's load, we have $\sum_{v,k}u_{vk}\le b-1$, so each $v_k$ has enough fixed representatives for these swaps.
To show that the swaps leave bin $v$ with load at most $T'_v$, fix a range $I_k$ containing movable representatives.
At any cost threshold, each upward update increases the movable mass above it by less than one, while downward updates cannot increase it.
Thus removing the $u_{vk}$ costliest movable representatives, or all if fewer remain, eliminates any excess above the initial mass at thresholds at least the minimum movable cost in $I_k$.
The incoming fixed representatives cost less than this minimum.
Below this minimum, the initial mass was $b_{vk}$, which bounds the final mass because the swaps preserve the number of representatives.
Hence the swaps leave no more mass above any threshold than in $z^*$, and therefore no larger cost.
Summing over $I_1,\ldots,I_d$, the load of each $v\notin\mathcal V$ is at most $T'_v\le T_v$.
The cycle updates decrease loads outside $\mathcal V$ by at most $\left(b-1\right)\cdot\kappa/d$ in total.
The at most $b-1$ swaps decrease these loads by at most the same amount, since each swap exchanges representatives from the same cost range.
Because total cost is conserved, the representatives now in $\mathcal V$ have total cost at most
\begin{equation*}
 \sum_{v\in\mathcal V}T'_v
 +2\left(b-1\right)\cdot\frac{\kappa}{d}
 \le \sum_{v\in\mathcal V}T'_v+2 d\cdot\kappa
 \le \sum_{v\in\mathcal V}T_v.
\end{equation*}
Pack these representatives in IDO order through the bins of $\mathcal V$, retaining each bin's original capacity $T_v$.
Whenever the next representative does not fit, exclude it and move to the next bin.
The total cost of the remaining representatives stays at most the total capacity of the remaining bins.
Therefore, all representatives fit by the final bin, and at most $\left|\mathcal V\right|-1$ further representatives are excluded.
Together with Step~1, at most $5 d-1=O(\sqrt L)$ additional representatives are excluded.

\paragraph{Step 5: Counting Segments.}
Before the swaps, list the movable representatives of each $I_k$ in cost order and record their assigned bins, suppressing consecutive repetitions.
The resulting sequence contains no alternating subsequence $u,v,u,v$, by the order property from Step~2.
In particular, each ordered pair of distinct bins can appear as consecutive labels at most once, so there are at most $b(b-1)+1$ segments per range.
The $d\cdot b\cdot\left(b-1\right)$ fixed representatives and at most $b-1$ swaps introduce only $O\!\left(d\cdot b^2\right)$ further boundaries.
The final repacking adds only $O(d)$ boundaries.
Including the $M_\ell+2 d$ representatives excluded before rounding, the total number of segments is
$O(d\cdot b^2+\sqrt L)=O\!\left(L^3\right)$.
Expanding each representative $\hat e_t$ into its contiguous block $B_t$ keeps the bin or exclusion label constant within that block.
Merging the fractional-capacity bin into unit-capacity bin $L$ cannot increase the number of segments.
Excluding the first $\min\left\{q+h,\left|G_{i,L}\right|\right\}$ chores of $G_{i,L}$ introduces at most one new boundary, so the base vectors $\left(y_{i,t}(e)\right)_t$ have $O\!\left(L^3\right)$ constant segments.
The order-preserving shift to $M\setminus H$ deletes the excluded positions in $Z_i$, keeps the first $m-K$ remaining positions, and closes the gaps; none of these operations increases the number of segments.
Assigning the constant vector $\left(\tau_{i,t}\right)_t$ to $H=\left\{e_1,\ldots,e_K\right\}$ adds at most one boundary.
Thus the vectors $\left(x_{i,t}(e)\right)_t$ change value only $O\!\left(L^3\right)$ times, completing the proof.

\subsection{Proof of Lemma~\ref{lem:reserve-coordination}}\label{app:joint-selection}


Throughout this proof, agent indices $i$ range over $N_o$.
Recall that
\begin{equation*}
 \sum_t\eta_{i,t}=1,
 \qquad
 0\le x_{i,t}(e)\le\frac1q,
 \qquad
 \sum_{i,t}\eta_{i,t}\cdot x_{i,t}(e)=\frac{n-k}{n}
 \qquad\forall e\in M.
\end{equation*}
Let $\Gamma$ denote the number of distinct vectors $\left(x_{i,t}(e)\right)_{i,t}$ as $e$ varies over $M$.
Fix a sufficiently large absolute constant $C_4>0$.
Let $\mathcal S$ be the set of all nonnegative coefficient vectors $\zeta$ satisfying $\sum_t\zeta_{i,t}=1$ for every $i\in N_o$, the $k$ reserved-agent cost equations
\begin{equation*}
 \sum_{i,t}\zeta_{i,t}\cdot c_j\left(x_{i,t}\right)
 =\left(n-k\right)\cdot\PROP_j
 \qquad\forall j\in N_r,
\end{equation*}
at most $2(k+1)$ mixed agents, and
\begin{equation*}
 \left|\sum_{i,t}\left(\zeta_{i,t}-\eta_{i,t}\right)\cdot x_{i,t}(e)\right|
 \le\frac{C_4}{2}\cdot
 \sqrt{\frac{\log\left(2\Gamma\right)}q}
 \qquad\forall e\in M.
\end{equation*}
The cost equations hold at $\eta$ because each ordinary agent has expected fraction $1/n$ of every chore.
We first show that, for every linear functional $f$, some $\zeta\in\mathcal S$ satisfies $f(\zeta)=f(\eta)$.
By the separating-hyperplane theorem, this implies that $\eta$ is a finite convex combination of points in $\mathcal S$, yielding a lottery with $\mathbb E\left[\widehat{\eta}\right]=\eta$.

\paragraph{Step 1:}
Fix $f$ and start Bansal's randomized rounding procedure~\cite[Theorem~1.2]{journals/toc/000124} from $\eta$, with parameter $1/4$.
Keep the agent sum equations, the $k$ reserved-agent cost equations, and $f(\zeta)=f(\eta)$ active.
When $F$ coordinates are strictly fractional, at most $F/2$ agents are mixed, since a mixed agent has at least two fractional coefficients.
Thus, while $F>4(k+1)$, the rank of the active equations on these coordinates is at most $\frac F2+k+1<\frac{3 F}{4}$, as required by Bansal's procedure.
Stop as soon as at most $4(k+1)$ coordinates remain fractional, and denote the resulting vector by $\zeta$.
It preserves all active equations and has at most $2(k+1)$ mixed agents.

\paragraph{Step 2:}
For each chore $e$, the coefficients of its allocated fraction satisfy
\begin{equation*}
 \sum_{i,t}x_{i,t}(e)^2\cdot\eta_{i,t}\cdot\left(1-\eta_{i,t}\right)
 \le\frac1q\cdot\sum_{i,t}\eta_{i,t}\cdot x_{i,t}(e)
 \le\frac1q
 \quad\text{and}\quad
 \max_{i,t}\left\{x_{i,t}(e)\right\}\le\frac1q.
\end{equation*}
Bansal's concentration bound remains valid when the procedure is stopped as soon as at most $4(k+1)$ coefficients remain fractional~\cite[Theorem~1.2 and Section~3.2.1]{journals/toc/000124}.
Indeed, freezing the process after it stops preserves the exponential supermartingale used in that proof.
Its remaining variance term is nonnegative, so dropping this term gives the same bound for the stopped linear sum.
Consequently, for every $u>0$,
\begin{equation*}
 \Pr\left[
 \left|\sum_{i,t}\left(\zeta_{i,t}-\eta_{i,t}\right)\cdot x_{i,t}(e)\right|
 \ge u\right]
 \le2\cdot\exp\left(-\frac{q\cdot u^2}
 {160\left(1+u/3\right)}\right).
\end{equation*}
By Appendix~\ref{app:rearranging}, the vector $\left(x_{i,t}(e)\right)_t$ changes value $O\!\left(L^3\right)$ times along the chore order for each ordinary agent.
Taking the union of these change points gives $\Gamma=O\!\left(n\cdot L^3\right)=O\!\left(n^4\right)$.
Thus $\log\left(2\Gamma\right)\le6\cdot\log n$ for sufficiently large $n$.
Since $q\ge4\cdot\log n$, we have $\log\left(2\Gamma\right)/q\le3/2$.
Taking $u=\frac{C_4}{2}\cdot
\sqrt{\frac{\log\left(2\Gamma\right)}q}$ therefore makes the tail probability at most $1/\left(2\Gamma\right)$ for sufficiently large $C_4$.
A union bound over the $\Gamma$ distinct vectors $\left(x_{i,t}(e)\right)_{i,t}$ shows that $\zeta\in\mathcal S$ with probability at least $1/2$.
Since $f(\zeta)=f(\eta)$ holds throughout the process, a suitable realization exists.

\paragraph{Step 3:}
The set $\mathcal S$ is compact: fixing the mixed agents and the selected bundle of each pure agent gives a closed subset of the product of the agents' coefficient simplices, and there are finitely many such choices.
Its convex hull is therefore compact as well.
If $\eta\notin\operatorname{conv}(\mathcal S)$, strict hyperplane separation would give a linear functional $f$ such that $f(\eta)>\max_{\zeta\in\mathcal S}\left\{f(\zeta)\right\}$, contrary to the realization found above.
Hence $\eta\in\operatorname{conv}(\mathcal S)$, and a finite convex combination expressing $\eta$ supplies a lottery over $\widehat{\eta}\in\mathcal S$ with $\mathbb E\left[\widehat{\eta}\right]=\eta$.
Finally, the bound on $\Gamma$ from Step~2 and the assumption $k\ge C_2\cdot n\cdot\sqrt{\left(\log n\right)/q}$ imply, for $C_2\ge8 C_4$, that
\begin{equation*}
 \frac{C_4}{2}\cdot
 \sqrt{\frac{\log\left(2\Gamma\right)}q}
 \le\frac{\sqrt6\cdot C_4}{2}\cdot\sqrt{\frac{\log n}q}
 \le\frac kn.
\end{equation*}
Thus every supported outcome satisfies
\begin{equation*}
 x(e)=\sum_{i,t}\widehat{\eta}_{i,t}\cdot x_{i,t}(e)
 \le\frac{n-k}{n}+\frac kn=1
 \qquad\forall e\in M.
\end{equation*}
The lottery preserves all reserved-agent cost equations and all marginal coefficients, and every supported outcome has at most $2(k+1)$ mixed agents, as required.

\section{AMMS and \texorpdfstring{$1$-out-of-$d$ MMS}{1-out-of-d MMS}}\label{app:amms-ordinal}

In this appendix, we adapt the framework of Section~3 to establish our remaining two guarantees: ex-ante proportionality combined with ex-post $\tilde{O}(n^{2/3})$-AMMS (where the non-MMS agent is chosen uniformly at random) and with ex-post $1$-out-of-$d$ MMS for $d=n-\tilde{O}(n^{2/3})$.
As in Section~3, we may assume without loss of generality that $n$ is sufficiently large, since for any bounded $n$, assigning all chores to a uniformly chosen agent is ex-ante proportional, $n$-AMMS, and $1$-out-of-$1$ MMS.

\subsection{A General Framework}\label{sec:varying-reserve}

In Section~3, packing the representative chores into $\lfloor n/q\rfloor$ unit-capacity bins required excluding a subset of chores $H$ and restoring fractions of $H$ into the ordinary agents' fractional bundles.
As a consequence, any ordinary agent who received a chore from $H$ after rounding could exceed her MMS bound and only be guaranteed MMS1.
For AMMS and $1$-out-of-$d$ MMS, however, we cannot allow any exceptions among the ordinary agents.
We therefore modify the three-step construction of Section~3 as follows:
\begin{itemize}
    \item First, instead of excluding a subset of chores, we use the asymptotic bin-packing bound of Hoberg and Rothvoss~\cite{conf/soda/HobergR17} to pack \emph{all} representative chores into a slightly larger number of unit-capacity bins $L$. Because the resulting fractional bundles are rounding-resilient and partition $M$, every pure ordinary agent is guaranteed to meet her target benchmark after unit-slot rounding.
    \item Second, we coordinate the ordinary agents' lotteries using the proof of Lemma~\ref{lem:reserve-coordination} without reserved-agent cost equations, and finish the selection for the remaining mixed agents so that every ordinary agent is pure.
    Since $q\cdot L>n-k$, the ordinary agents together receive an expected fraction $\left(n-k\right)/\left(q\cdot L\right)<1$ of every chore, leaving slack for the concentration argument.
    The remaining unassigned fractions are rounded via general unit-slots into a residual chore set $R$.
    \item Third, we allocate the residual chore set $R$ among the $k$ reserved agents in $N_r$ so that every $j\in N_r$ incurs conditional expected cost at most $c_j(R)/k$.
    Over the uniformly random choice of $N_r$, this bound ensures ex-ante proportionality for every agent.
\end{itemize}

Fix an integer $1\le k<n$, where we will set $k=1$ for AMMS and $k=n-d=\tilde{O}(n^{2/3})$ for $1$-out-of-$d$ MMS.
We then sample a $k$-element set of reserved agents $N_r\subseteq N$ uniformly at random and let $N_o=N\setminus N_r$ denote the set of $n-k$ ordinary agents.
By scaling each agent's cost function $c_i$, we assume without loss of generality that the target benchmark equals $1$ for every agent $i\in N$ (namely, $\mu_i=1$ for AMMS and $\operatorname{MMS}_i^d(M)=1$ with $d=n-k$ for $1$-out-of-$d$ MMS).
In either case, each agent $i$ admits a partition of $M$ into $n-k+1$ bundles of cost at most $1$ (by adding an empty bundle when $d=n-k$), which implies
\begin{equation*}
 c_i(M)\le n-k+1 \qquad \text{and} \qquad \kappa_i\le 1 \qquad \forall i\in N.
\end{equation*}
We set the block size $q$ and the number of fractional bundles $L$ as
\begin{equation*}
 q=\left\lfloor\left(\frac{n^2}{\log n}\right)^{1/3}\right\rfloor
 \qquad \text{and} \qquad
 L=\left\lceil\frac{n-k}{q}+C_5\cdot\log n\right\rceil,
\end{equation*}
where $C_5>0$ is a sufficiently large absolute constant.
Thus $q=\tilde{O}(n^{2/3})$ and $L=\tilde{O}(n^{1/3})$.

\paragraph{Lottery of Fractional Bundles.}
As in Section~\ref{sec:fractional-allocations}, we assume by adding zero-cost dummy chores that $m$ is a multiple of $q$, and partition $M$ into $m/q$ blocks $B_1,\ldots,B_{m/q}$ of size $q$ with representative chores $\hat e_1,\ldots,\hat e_{m/q}$, where $\hat e_j=e_{(j-1)q+1}$.
Instead of excluding $B_1$ and $O(\sqrt{L})$ additional blocks as in Lemma~\ref{lem:core-reduction}, we pack all representative chores into $L$ unit-capacity bins.

The following lemma shows that, for every agent, we can partition the representative sequence into $O(L^2)$ contiguous segments and pack these segments into $L$ bins of cost at most $1$ so that every segment is placed entirely within a single bin.

\begin{lemma}\label{lem:complete-packing-source}
For each agent $i\in N$, all representative chores $\hat e_1,\ldots,\hat e_{m/q}$ can be partitioned into $L$ bins of cost at most $1$, such that the bin assignment changes at most $O(L^2)$ times along $\hat e_1,\ldots,\hat e_{m/q}$.
\end{lemma}
\begin{proof}
Fix an agent $i\in N$ and a partition of $M$ into $n-k+1$ bundles of cost at most $1$.
As in Appendix~\ref{app:repeated-halving}, replacing each chore in $B_\ell$ ($\ell<m/q$) by a copy of $\hat e_{\ell+1}$ and discarding the chores in $B_{m/q}$ packs $q$ copies of each representative $\hat e_2,\ldots,\hat e_{m/q}$ into $n-k+1$ unit-capacity bundles.
Assigning weight $1/q$ to each of these $n-k+1$ bundles and adding the singleton bundle $\{\hat e_1\}$ with weight $1$ produces a fractional bin packing of $\hat e_1,\ldots,\hat e_{m/q}$ of total weight $(n-k+1)/q+1\le L-1$.
In particular, $\sum_{j=1}^{m/q}c_i(\hat e_j)\le L-1$.

We call a representative $\hat e_j$ \emph{large} if $c_i(\hat e_j)\ge 1/L$, and \emph{small} otherwise.
Since $\sum_{j=1}^{m/q}c_i(\hat e_j)<L$, the large representatives form a prefix of fewer than $L^2$ items.
When restricted to the large representatives, the fractional packing above yields a fractional configuration packing of value at most $(n-k+1)/q+1$.
By the asymptotic bin-packing theorem of Hoberg and Rothvoss~\cite[Theorem~2]{conf/soda/HobergR17}, the large representatives can be packed integrally into at most $(n-k+1)/q+1+O(\log n)$ unit-capacity bins, and this number of bins is at most $L$ for a sufficiently large absolute constant $C_5>0$.
The remaining small representatives form a contiguous suffix of $\hat e_1,\ldots,\hat e_{m/q}$.
We pack them in IDO order into the remaining capacities of the $L$ bins by moving to the next bin whenever the next small representative does not fit in the current bin.
Whenever a bin is closed during this pass, its load strictly exceeds $1-1/L=(L-1)/L$.
Since the total cost of all representatives is at most $L-1$, at most $L-1$ bins can be closed, so all small representatives fit into the $L$ bins.
Finally, the prefix of large representatives has fewer than $L^2$ chores and the greedy pass partitions the suffix of small representatives into at most $L$ contiguous intervals, so the bin assignment changes at most $O(L^2)$ times along $\hat e_1,\ldots,\hat e_{m/q}$.
\end{proof}

For each ordinary agent $i\in N_o$ and each bin $t\in[L]$ from Lemma~\ref{lem:complete-packing-source}, let $G_{i,t}$ be the union of the blocks $B_j$ whose representatives $\hat e_j$ belong to bin $t$.
We define the fractional bundles and their coefficients by
\begin{equation*}
 x_{i,t}(e)=\frac1q\cdot\mathbf{1}\left(e\in G_{i,t}\right)
 \qquad \text{and} \qquad
 \eta_{i,t}=\frac1L
 \qquad \forall t\in[L].
\end{equation*}
Since $G_{i,1},\ldots,G_{i,L}$ partition $M$, every chore $e\in M$ has expected assigned fraction $\sum_{t=1}^L\eta_{i,t}\cdot x_{i,t}(e)=\frac1{q\cdot L}$.
We place the $q$ fractions of size $1/q$ from every block $B_j\subseteq G_{i,t}$ into a single private slot, which ensures that $x_{i,t}$ is rounding-resilient.
Specifically, any integral bundle that contains at most one chore from every slot of $x_{i,t}$ has cost at most $\sum_{\hat e_j\text{ in bin }t}c_i(\hat e_j)\le 1$.

\paragraph{Coordinating the Lotteries.}
For each ordinary agent $i\in N_o$, the vector $(x_{i,t}(e))_t$ is constant on each block $B_j$ and changes value at most $O(L^2)$ times along $e_1,\ldots,e_m$.
To apply the concentration argument, we need a common partition of the chore order on which every ordinary agent's candidate fractions are constant.
Taking the union of the segment boundaries across all $n-k$ ordinary agents yields $O(n\cdot L^2)$ consecutive segments with this property.
As in Appendix~\ref{app:joint-selection}, $\Gamma$ denotes the number of distinct vectors $(x_{i,t}(e))_{i,t}$ as $e$ ranges over $M$.
Thus, $\Gamma=O(n\cdot L^2)=O(n^3)$, so $\Gamma\le n^5$ and $\log(2\Gamma)\le 6\log n$ for sufficiently large $n$.

We use the proof of Lemma~\ref{lem:reserve-coordination} in Appendix~\ref{app:joint-selection} without the reserved-agent cost equations.
Here the initial total fraction of every chore assigned to ordinary agents is $\left(n-k\right)/\left(q\cdot L\right)\le1$, so the same variance bound and concentration argument apply.
Recall that $F$ denotes the number of coefficients strictly between $0$ and $1$ at the current stage of the rounding procedure.
Note that every mixed agent has at least two such coefficients, so at most $F/2$ agent sum equations involve these coordinates.
The auxiliary equation for $f$ increases the rank by at most one.
Thus, the rank on the fractional coordinates is at most $F/2+1<3\cdot F/4$ whenever $F>4$.
Stopping when at most four coordinates remain fractional therefore leaves at most two mixed agents.
The same convexity argument gives a finite lottery over coefficient vectors $\zeta$ with $\mathbb E[\zeta]=\eta$, at most two mixed agents, and discrepancy at most $\frac{C_4}{2}\cdot\sqrt{\log\left(2\Gamma\right)/q}$ for every chore.

Conditional on $\zeta$, independently select one bundle for each mixed agent $i$, choosing bundle $t$ with probability $\zeta_{i,t}$, and keep each pure agent's selected bundle.
The resulting binary coefficient vector $\widehat{\eta}$ satisfies $\mathbb E\left[\widehat{\eta}\mid\zeta\right]=\zeta$, and hence $\mathbb E\left[\widehat{\eta}\right]=\eta$.
Thus every ordinary agent is pure and receives a single prepared bundle $x_i=x_{i,t}$.
Each mixed agent's fraction of a chore lies in $\left[0,1/q\right]$ both before and after this selection.
Since there are at most two mixed agents, the total assigned fraction of any chore changes by at most $2/q$.
With $x(e)=\sum_{i\in N_o,t}\widehat{\eta}_{i,t}\cdot x_{i,t}(e)$, we therefore have, in every supported outcome,
\begin{equation*}
 \left|x(e)-\frac{n-k}{q\cdot L}\right|
 \le\frac{C_4}{2}\cdot
 \sqrt{\frac{\log\left(2\Gamma\right)}q}
 +\frac2q
 \le6\cdot C_4\cdot\sqrt{\frac{\log n}q}
 \qquad\forall e\in M,
\end{equation*}
where the first inequality combines the discrepancy bound for $\zeta$ with the additional $2/q$ from the final selection.
The second uses $\log(2\Gamma)\le6\log n$ and $2/q\le C_4\sqrt{(\log n)/q}$ for sufficiently large $n$.
We next show that the unassigned fraction under the initial coefficients is large enough to absorb this discrepancy.
The definition of $L$ gives
\begin{equation*}
 n-k+C_5\cdot q\log n\le q\cdot L
 <n-k+C_5\cdot q\log n+q\le2n,
\end{equation*}
where the last inequality holds for sufficiently large $n$ because $q\log n=o(n)$.
By our choice of $q$, both $q\log n$ and $n\sqrt{(\log n)/q}$ are $\Theta\!\left(n^{2/3}(\log n)^{2/3}\right)$.
Hence, for sufficiently large $C_5$ and $n$,
\begin{equation*}
 6C_4\cdot\sqrt{\frac{\log n}{q}}
 \le\frac{C_5\cdot q\log n}{2n}
 \le\frac{C_5\cdot q\log n}{q\cdot L},
\end{equation*}
where the second inequality uses $q\cdot L\le2n$.
It follows that
\begin{equation*}
 x(e)\le\frac{n-k}{q\cdot L}+\frac{C_5\cdot q\log n}{q\cdot L}\le 1 \qquad \forall e\in M.
\end{equation*}

\paragraph{Rounding Fractional Allocations.}
We organize each ordinary agent's bundle $x_i$ ($i\in N_o$) into its block-slots defined above, and organize the remaining unassigned fractions $1-x(e)$ in IDO order into general unit-slots as in Section~\ref{sec:rounding}.
Rounding all slots jointly via Lemma~\ref{lem:slots} assigns each ordinary agent $i\in N_o$ an integral bundle $X_i$ of cost $c_i(X_i)\le 1$, while the chores matched to the remaining slots form the residual chore set $R=M\setminus\bigcup_{i\in N_o}X_i$.
Because the remaining fractions $1-x(e)$ are organized into general unit-slots in the common IDO order, the unit-slot bound of Lemma~\ref{lem:slots} applies simultaneously to every agent $i\in N$.
For every chore $e\in M$, the discrepancy bound above gives 
$$
1-x(e)\le\frac{q\cdot L-n+k}{q\cdot L}+6C_4\cdot\sqrt{(\log n)/q}.
$$
For every agent $i\in N$, we therefore have
\begin{equation}\label{eq:scaled-residual-estimates}
 \begin{aligned}
 c_i(R)
 &\le\sum_{e\in M}c_i(e)\cdot\left(1-x(e)\right)+\kappa_i \le\left(\frac{q\cdot L-n+k}{q\cdot L}
      +6C_4\cdot\sqrt{\frac{\log n}{q}}\right) \cdot c_i(M)+1\\
 &\le q\cdot L-n+k+6C_4\cdot n\sqrt{\frac{\log n}{q}}+1=O\!\left(n^{2/3}(\log n)^{2/3}\right).
 \end{aligned}
\end{equation}
The first inequality is the general unit-slot bound, and the second uses the upper bound on $1-x(e)$ together with $\kappa_i\le1$.
In the third inequality, we use $c_i(M)\le q\cdot L$ for the first term and $c_i(M)\le n-k+1\le n$ for the second.
Finally, the definition of $L$ gives $q\cdot L-n+k\le C_5\cdot q\log n+q$, so the last bound follows from our choice of $q$.
Its implied constant is independent of $k$.
Furthermore, conditional on $N_r$, Lemma~\ref{lem:slots} preserves the marginal probability of every chore: each ordinary agent $i\in N_o$ receives $e\in M$ with probability $\mathbb E\left[x_i(e)\mid N_r\right]=\frac1{q\cdot L}$, and $e\in R$ with probability $\mathbb E\left[1-x(e)\mid N_r\right]=1-\frac{n-k}{q\cdot L}$.
Consequently, if we allocate $R$ among the $k$ reserved agents so that each $i\in N_r$ incurs conditional expected cost at most $c_i(R)/k$, then averaging over the uniformly chosen $k$-element subset $N_r\subseteq N$ yields
\begin{equation}\label{eq:reserve-average}
 \mathbb E\!\left[c_i(X_i)\right]
 \le\frac{n-k}{n}\cdot\frac{c_i(M)}{q\cdot L}
 +\frac{k}{n}\cdot\frac1k\left(1-\frac{n-k}{q\cdot L}\right)c_i(M)
 =\frac{c_i(M)}{n}=\PROP_i
 \qquad \forall i\in N.
\end{equation}

\subsection{Ex-ante PROP and Sublinear AMMS}\label{sec:amms}

To obtain our $\tilde{O}(n^{2/3})$-AMMS guarantee, we set $k=1$ and assign the entire residual chore set $R$ to the single reserved agent.

\begin{theorem}\label{thm:amms}
Every additive chores instance admits a finite ex-ante proportional lottery of $\tilde{O}(n^{2/3})$-AMMS allocations, with the non-MMS agent chosen uniformly at random.
\end{theorem}
\begin{proof}
Apply the framework of Section~\ref{sec:varying-reserve} with $k=1$ under the normalization $\mu_i=1$, and assign $X_j=R$ to the uniformly chosen reserved agent $N_r=\{j\}$.
Every ordinary agent $i\in N\setminus\{j\}$ receives a bundle of cost $c_i(X_i)\le 1$, while~\eqref{eq:scaled-residual-estimates} bounds the reserved agent's cost by $c_j(X_j)=c_j(R)=\tilde{O}(n^{2/3})$.
Since $k=1$, agent $j$ trivially receives cost $c_j(R)/k$, so~\eqref{eq:reserve-average} implies $\mathbb E\!\left[c_i(X_i)\right]\le\PROP_i$ for all $i\in N$.
Note that both guarantees can transfer to the original instance by Lemma~\ref{lem:ido-reduction}.
\end{proof}

\subsection{Ex-ante PROP and \texorpdfstring{$1$-out-of-$d$ MMS}{1-out-of-d MMS}}\label{sec:ordinal}

To obtain $1$-out-of-$d$ MMS for every agent, we choose $k=\tilde{O}(n^{2/3})$ large enough that~\eqref{eq:scaled-residual-estimates} guarantees $c_i(R)\le k/2$ for every agent $i\in N$, and set $d=n-k$.
We then allocate $R$ among the $k$ reserved agents using Algorithm~\ref{alg:bundle-completion} and unit-slot rounding from Section~\ref{sec:four-thirds}.

By adding zero-cost dummy chores to $R$ if necessary, we may assume that $|R|\ge k$.
We take the $k$ costliest chores of $R$ under the IDO ordering and place them into singleton bundles $Q_1,\ldots,Q_k$, with one chore per bundle.
The remaining chores form $V=R\setminus\bigcup_{j=1}^k Q_j$.
For every reserved agent $i\in N_r$, each singleton bundle has cost $c_i(Q_j)\le\kappa_i\le 1$.
Moreover, since $Q_1,\ldots,Q_k$ and any chore $e\in V$ form $k+1$ chores in $R$ that each cost at least $c_i(e)$, we have
\begin{equation*}
 \max_{j\in[k]}\left\{c_i(Q_j)\right\}\le 1
 \qquad \text{and} \qquad
 \max_{e\in V}\left\{c_i(e)\right\}\le\frac{c_i(R)}{k+1}\le\frac{c_i(R)}{k}.
\end{equation*}
The completion procedure and Lemmas~\ref{lem:fractional-completion} and~\ref{lem:bundle-completion} extend to any partition into fixed bundles that are MMS1 for every participating agent, together with the remaining chores.
Indeed, the fractional completion proof uses only the partition property, and the rounding argument requires the fixed bundles to be MMS1 only when an agent receives no additional chores.
Here every fixed bundle $Q_j$ is a singleton and hence is MMS1 for every reserved agent.
Applying Algorithm~\ref{alg:bundle-completion} and Lemma~\ref{lem:bundle-completion} to the $k$-agent instance $(N_r,R)$ with fixed bundles $Q_1,\ldots,Q_k$ and remaining chores $V$ ensures that, conditional on $N_r$ and $R$, each reserved agent $i\in N_r$ has expected cost at most $c_i(R)/k$.
Furthermore, in every supported outcome, either agent $i\in N_r$ receives only her singleton bundle $Q^i$ (of cost $c_i(Q^i)\le 1$), or there exists a chore $e\in X_i\cap V$ such that $c_i(X_i\setminus\{e\})\le c_i(R)/k$, which yields
\begin{equation*}
 c_i(X_i)\le\max\left\{1,\frac{c_i(R)}{k}+\max_{e\in V}\left\{c_i(e)\right\}\right\}
 \le\max\left\{1,\frac{2c_i(R)}{k}\right\},
\end{equation*}
after which we discard any added zero-cost dummy chores.

\begin{theorem}\label{thm:ordinal}
Every instance of additive chores admits a finite ex-ante proportional lottery in which every allocation is $1$-out-of-$d$ MMS for $d=n-\tilde{O}(n^{2/3})$.
\end{theorem}
\begin{proof}
By~\eqref{eq:scaled-residual-estimates}, we may choose $k=\tilde{O}(n^{2/3})$ so that $c_i(R)\le k/2$ for every agent $i\in N$ under the normalization $\operatorname{MMS}_i^d(M)=1$ with $d=n-k$.
Every ordinary agent $i\in N_o$ receives a bundle of cost $c_i(X_i)\le 1$.
The singleton bundle-completion procedure above ensures that each reserved agent $i\in N_r$ has conditional expected cost at most $c_i(R)/k$ and ex-post cost at most $\max\{1,2c_i(R)/k\}\le 1$.
By~\eqref{eq:reserve-average}, the overall lottery is ex-ante proportional, and every agent receives cost at most $1=\operatorname{MMS}_i^d(M)$ in every supported outcome.
Applying Lemma~\ref{lem:ido-reduction} transfers both guarantees to the original instance.
\end{proof}

\end{document}